\documentclass[preprint]{emulateapj}
\slugcomment{Accepted for Publication in the Astrophysical Journal}
\shortauthors{Zaritsky, et al.}
\shorttitle{}

\begin{document}
\title{Globular Cluster Populations: First Results from S$^4$G Early-Type Galaxies}
  
\author{Dennis Zaritsky\altaffilmark{1}, Manuel Aravena\altaffilmark{2}, E. Athanassoula\altaffilmark{3}, Albert Bosma\altaffilmark{3}, S\'ebastien Comer\'on\altaffilmark{4,5}, Bruce G. Elmegreen\altaffilmark{6}, Santiago Erroz-Ferrer\altaffilmark{7,8}, Dimitri A. Gadotti\altaffilmark{9}, 
Joannah L. Hinz\altaffilmark{11}, Luis C. Ho\altaffilmark{12,13}, Benne Holwerda\altaffilmark{14}, Johan H. Knapen\altaffilmark{7,8}, 
Jarkko Laine\altaffilmark{4}, 
Eija Laurikainen\altaffilmark{4,5}, 
Juan Carlos Mu\~noz-Mateos\altaffilmark{9,15}, Heikki Salo\altaffilmark{4},  and Kartik Sheth\altaffilmark{15}}

\altaffiltext{1}{Steward Observatory, University of Arizona, 933 North Cherry Avenue, Tucson, AZ 85721, USA; dennis.zaritsky@gmail.com}
\altaffiltext{2}{N\'ucleo de Astronom\'{\i}a, Facultad de Ingenier\'{\i}a, Universidad Diego Portales, Av. Ej\'ercito 441, Santiago, Chile}
\altaffiltext{3}{Aix Marseille Universit\'e, CNRS, LAM (Laboratoire d'Astrophysique de Marseille) UMR 7326, 13388, Marseille, France}
\altaffiltext{4}{Astronomy Division, Department of Physics, P.O. Box 3000, FI-90014 University of Oulu, Finland}
\altaffiltext{5}{Finnish Centre of Astronomy with ESO (FINCA), University of Turku, V\"ais\"al\"antie 20, FI-21500, Piikki\"o, Finland}
\altaffiltext{6}{IBM T. J. Watson Research Center, 1101 Kitchawan Road, Yorktown Heights, NY 10598, USA}
\altaffiltext{7}{Instituto de Astrof\'isica de Canarias, V\'{i}a L\'{a}cteas 38205 La Laguna, Spain}
\altaffiltext{8}{Departamento de Astrof\'{\i}sica, Universidad de La Laguna, 38206, La Laguna,  Spain}
\altaffiltext{9}{European Southern Observatory, Casilla 19001, Santiago 19, Chile}
\altaffiltext{10}{Departamento de Astrof\'{\i}sica y CC$.$ de la Atm\'osfera, Universidad Complutense de Madrid, Madrid, Spain }
\altaffiltext{11}{MMT Observatory, P.O. Box 210065, Tucson, AZ 85721, USA}
\altaffiltext{12}{Kavli Institute for Astronomy and Astrophysics, Peking University, Beijing 100871, China}
\altaffiltext{13}{Department of Astronomy, Peking University, Beijing 100871,China}
\altaffiltext{14}{University of Leiden, Leiden Observatory, Niels Bohrweg 4, NL-2333, Leiden, The Netherlands}
\altaffiltext{15}{National Radio Astronomy Observatory/ NAASC, 520 Edgemont Road, Charlottesville, VA 22903, USA }

\begin{abstract} 
Using 3.6$\mu$m images of 97 early-type galaxies, we develop and verify methodology to measure globular cluster populations from the S$^4$G survey images. We find that 1) the ratio, T$_{\rm N}$, of the number of clusters, N$_{\rm CL}$, to parent galaxy stellar mass, M$_*$, rises weakly with M$_*$ for early-type galaxies with M$_* > 10^{10}$ M$_\odot$ when we calculate galaxy masses using a universal stellar initial mass function (IMF), but that the dependence of T$_{\rm N}$ on M$_*$ is removed entirely once we correct for the recently uncovered systematic variation of IMF with M$_*$, and 2) for M$_* < 10^{10}$ M$_\odot$ there is no trend between N$_{\rm CL}$ and M$_*$, the scatter in T$_{\rm N}$ is significantly larger (approaching 2 orders of magnitude), and there is evidence to support a previous, independent suggestion of two families of galaxies. The behavior of N$_{\rm CL}$ in the lower mass systems is more difficult to measure because these systems are inherently cluster poor, but our results may add to previous evidence that large variations in cluster formation and destruction efficiencies are to be found among low mass galaxies. The average fraction of stellar mass in clusters is $\sim$ 0.0014 for M$_* > 10^{10}$ M$_\odot$ and can be as large as $\sim 0.02$ for less massive galaxies.
These are the first results from the S$^4$G sample of galaxies, and will be enhanced by the sample of early-type galaxies now being added to S$^4$G and complemented by the study of later type galaxies within S$^4$G.

\end{abstract}


\section{Introduction}
\label{sec:intro}
There are a few well-established empirical findings that help guide our developing understanding of galaxy formation and evolution. Among those are some that hint at the importance of environmental processes \citep{dressler80,pg}, often referred to as the role of nurture in galaxy evolution,  and others that highlight the role of halo mass \citep{kauffmann,bundy}, often referred to as the role of nature. Although numerous studies have attempted to establish the supremacy of one set of influences over another, the two are intertwined due to the relationship in hierarchical growth models between environment and mass \citep{delucia}.  

At the core of the hierarchical growth paradigm is the agglomeration of mass in galaxies, by mergers or accretion. Unfortunately, observational markers of the most significant events, such as tidal tails or bridges that would evince a recent major interaction or merger, are generally ambiguous and difficult to establish for
large samples \citep[see][for examples]{tal,lotz,adams}, while minor events are nearly imperceptible.

What happens to a galaxy as it merges, accretes smaller neighbors, and grows? The scrambling of stars, gas, and dark matter that occurs during such events erases traces of the progenitors and of the processes that occurred prior to and during the event. Much of our intuition is instead guided by simulations, and those are in critical need of empirical confirmation. Although there are specific examples where we catch a galaxy relatively soon after a burst of star formation, E+A or K+A galaxies \citep{dressler,couch} and where convincing evidence of a merger exists \citep{zab,norton,yang}, these are rare objects in the local universe \citep{zab}. For the more common, less dramatic accretion/merger events there is little in the way of tracers of the hierarchical accretion that is such an integral part of our understanding of the growth of structure. We need to identify a population of objects within galaxies that can testify to the combination of different progenitors and to any new star formation that occurs as part of the growth process.

An understanding of globular cluster populations in galaxies has developed over the past two decades that suggests that clusters could be such a tracer and provide new insights and constraints for modelers. There is empirical evidence that all but the least massive galaxies have multiple populations of globular clusters, where metallicity is the distinguishing characteristic  among the populations \citep[for a review see][]{brodie}, although alternative interpretations exist \citep{richtler,yoon} and the observational evidence is more complex than previously appreciated \citep{chies}. The developing interpretive consensus is that these populations reflect both an early epoch of cluster formation and subsequent evolution influenced by major dissipational mergers \citep{harris81,schweizer,ashman92}, dissipationless accretion \citep{cote}, and ongoing cluster formation \citep{kravtsov}. With increasingly sophisticated spectroscopic surveys \citep[such as that of][]{sluggs}, the connection between cluster and stellar populations will be tested and used to constrain formation models. 

Despite the promise of this field of study, {compiling large, homogeneous samples has been challenging}. {At a minimum, we need to understand whether}, the number of clusters per galaxy, normalized in some sensible way, varies with galaxy mass, morphology, environment, or stellar population. There is a long history of studies of the cluster number normalized by galaxy luminosity, a quantity defined as the specific frequency of clusters \citep[S$_{\rm N}$, see][for reviews]{harris,brodie}, and a more recent focus cluster number normalized by galaxy stellar mass \citep[T$_{\rm N}$;][]{zepf}, uncertainties due to contamination and completeness corrections were often large and galaxy samples were limited. The state-of-the-art compilation is that presented by \cite{harris13}, which scoured the literature to obtain such estimates for 422 galaxies, of which 341 are the early-types that are the focus here. Despite the care taken in that work, the one potential weakness of literature compilations is the unavoidable heterogeneity in sample selection, image quality and characteristics, image analysis, and cluster population modeling necessary for completeness corrections. This heterogeneity is particularly worrisome for identifying trends across a broad range of galaxy parameters because individual studies have tended to focus on a particular class of galaxies (for example, primarily galaxies in the Virgo cluster or of low-mass). As we describe below, there are a number of decisions involved in undertaking a cluster census and, in a relative sense, one is on surer footing if the data quality and analysis are the same across the entire sample. Therefore, a study such as ours at the very least constitutes a complementary approach with which to address these important questions.

Over the last ten years or so terrific progress has been made in obtaining high fidelity samples of clusters on which to base the cluster census, using both high angular resolution images, particularly those provided by the {\sl Hubble Space Telescope}, and color information from deep photometry, to remove contaminants \citep[for examples, see][]{peng06,strader,kundua,kundub,rhode04,young}. However, with a few exceptions \citep{peng06,strader}, these studies only cover a handful of galaxies because of the observational cost of obtaining such data. Those studies that have both high angular resolution and are photometrically sensitive tend to suffer from small fields of view that do not necessarily cover the full radial extent of the globular cluster population, creating yet another source of uncertainty in the final cluster counts \citep{rhode03}, and generally cover galaxies in a single environment \citep[such as the Virgo cluster;][]{peng06,strader}.

We have chosen to measure T$_{\rm N}$ for a large sample of galaxies in a manner that is more reminiscent of earlier treatments. The recent emphasis has been on greater and greater precision in the removal of contaminants. That focus has been in large part driven by the need for pure samples on which to base subsequent photometric \citep{gebhardt, larsen, spitler08a} and spectroscopic studies \citep{sluggs}. By relaxing the criteria on sample purity, because we are not seeking to define samples for follow-up studies, we can accept larger uncertainties in the measurement of the number of clusters, N$_{\rm CL}$, in exchange for larger galaxy samples that span a greater range of galaxy properties. 
If, as expected, N$_{\rm CL}$ varies by several orders of magnitude among galaxies, then uncertainties as large as a factor of a few may have little effect on broad trends. Ultimately, the final answer to whether such lower precision is scientifically useful depends on the magnitude of the effects present and the size of the sample. A basic weaknesses of our approach is that we treat the cluster population as a single entity, despite clear evidence from colors \citep{zepf,ostrov,gebhardt,larsen,kundua} and kinematics \citep{strader,woodley,pota} that there are multiple populations. Of course, a similar criticism can be levied on studies of stellar populations and yet basic relationships are still valid for the whole.

We proceed along these lines in an attempt to eventually obtain the largest, homogenous sample of cluster population measurements with which to complement the more intensive, focused work that is ongoing on smaller samples and the broader work being carried out with compilations of published results.
Our measurement of the cluster populations is based on a statistical excess of point sources in the S$^4$G images \citep{sheth} of nearby galaxies. 
We will establish that the methodology presented here is sufficiently accurate and precise to be scientifically useful.
The S$^4$G data have several direct advantages because imaging at IR wavelengths suffers less extinction than that at optical wavelengths and the IR luminosity is closely tied to older stellar populations. In addition, the S$^4$G data is a roughly volume limited sample of several thousand galaxies. Indirect advantages accrue from ancillary studies that include photometric decomposition \citep{kim,munoz,salo}, detailed and homogeneous morphological classification \citep{elmegreen,holwerda,buta14}, radial and vertical disk structures \citep{jlaine,comeron}, classification of asymmetric structures \citep{zlop,laine}, and stellar and gaseous mass estimates \citep{zbtf,querejeta}. In \S 2 we describe the sample, how we constructed the cluster candidate catalog, and how we measured the number of clusters and their radial distributions. We discuss the findings in \S3 and conclude in \S4. 

\section{The Data and Measurements}
\label{sec:data}

\subsection{Constructing the Point Source Catalog and Surface Densities}

The parent sample for this study is the S$^4$G sample, which currently consists of 2,352 galaxies selected as described by \cite{sheth}. Although primarily a volume-limited sample, some additional selection criteria,  such as the existence of an H{\small I} redshift, and the surface brightness limitations of the existing catalogs from which the sample was selected, preclude it from being a complete, volume-limited sample. We observed these galaxies using the {\sl Spitzer Space Telescope} \citep{werner} and its Infrared Array Camera \citep[IRAC][]{irac} as described by \cite{sheth}.  The data from the original S$^4$G survey are publicly available through the NASA IRSA website\footnote{http://irsa.ipac.caltech.edu/data/SPITZER/S4G/}. An extension of S$^4$G to cover the missing H{\small I}-weak galaxies in this volume that otherwise match our selection criteria was approved for Cycle 10 and {is expand to increase several fold} the number of galaxies relevant for this study, which is why we consider the results presented here to be the first of the set of results to come from S$^4$G regarding globular cluster populations in early type galaxies.

From among the completed original sample (DR1), we have so far limited our study. We focus on galaxies with morphologies identified as early type ($-5 \le$ T-type $ \le 1$) by \cite{buta14}, to avoid 
those galaxies with significant internal structure that could be mistaken for point sources. All but five in our sample have $T \le -2$. This morphological cut results in the sample for this study being a small fraction of the full S$^4$G sample. We further constrain the sample by including only those galaxies within a suitable range of distances. We set the upper end of the distance range to correspond to a distance modulus of 32.4 (30.1 Mpc), which we found to be the upper limit at which we have enough physical resolution within the central galaxy to result in a reliable globular cluster population radial profile (see below for more discussion of radial profile limits). We set the lower end of the distance modulus range at 30.25 (11.2 Mpc) which we found to be necessary to ensure sufficient background coverage within the images with which to constrain the background source density. 
We use redshift independent distances when available in NED, otherwise we use the redshift and adopt H$_0 = 72$ km s$^{-1}$ Mpc$^{-1}$ to derive a Hubble 
velocity distance estimate.

For this range of distances, the {\sl Spitzer} images are of sufficiently large angular size to reach well into background source population --- as defined by a lack of a detectable point source surface density gradient ---  but they often contain a significant number of bright stars, and occasionally other large galaxies. To be specific, S$^4$G images cover a projected scale of at least 1.5D$_{25}$, where D$_{25}$ is the diameter of the galaxy's isophote at the surface brightness level of 25 mag arcsec$^{-2}$, except for a subset of 125 archival galaxies. In terms of physical separations, we probe to galactocentric projected radii $>$ 30 kpc in all but one case, and typically out to between 50 and 100 kpc. Although there are certainly some clusters beyond these radii in many of our galaxies, we will show in \S2.3 that any uncertainties introduced by the lack of coverage at larger radii is subdominant to other uncertainties. All bright objects are masked in the critical parts of the analysis using the masks developed as part of the S$^4$G processing \citep{salo,munoz}. We remove from our sample those galaxies where another comparably large galaxy is included within the 3.6$\mu$m frame, both because of the increased difficulty in doing the measurement but also because interpreting the distribution of candidate globular clusters is more complicated. In a few other cases, we exclude the galaxy for technical reasons. Our final sample of 97 galaxies is given in Table \ref{tab:results}.

\begin{figure*}[t]
\plottwo{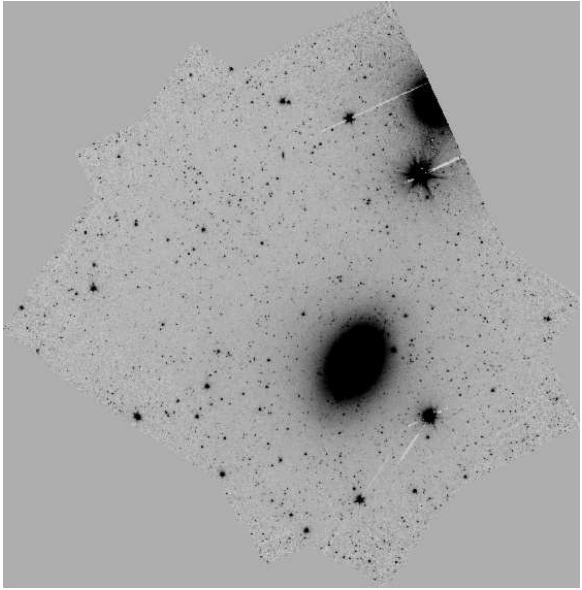}{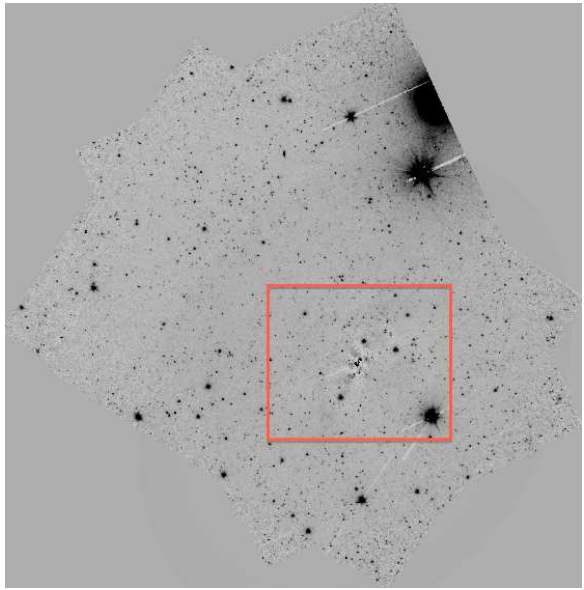}
\caption{Demonstration of the model subtraction for NGC 1553. We selected this galaxy as an example of a fairly luminous, extended galaxy in our sample. The results are otherwise not unusual. This is the S$^4$G produced mosaic in the 3.6$\mu$m passband. The size of the images is  24.5$^\prime$ across and north is up. The box in the right panel marks the area displayed in Figure \ref{fig:modelzoom}}.
\label{fig:modelsub}
\end{figure*}

Globular cluster candidates at these distances, in images of this angular resolution, will appear as point sources. It is therefore not possible based solely on morphology to separate foreground stars, galaxies of small angular size, and globular clusters. Our measurement will be statistical in nature, based on the expectation that contaminants will not cluster about the galaxy. Although colors, in principle, can help in distinguishing between these classes, in practice (see below) neither the other available {\sl Spitzer} band at 4.5$\mu$m, nor the available SDSS optical imaging, is of use here. Therefore, we are unable to identify any specific individual sources as likely clusters, but only provide an estimate of the excess of 3.6$\mu$m sources correlated with each galaxy and the radial distribution of this excess. We will argue that this excess is due to the globular cluster population and, where possible, validate this claim with comparison to published measurements. However, these characteristics preclude the use of these data, if not complemented with other bands, in studies of cluster subpopulations or for spectroscopic target selection.

Before cataloging all point sources with apparent magnitudes that are consistent with those of clusters at the corresponding distance, we process the images to aid us in identifying such objects. 
In addition to applying the masks mentioned previously (masked regions are not considered in the subsequent analysis other than in corrections for completeness), we use the exposure weight maps to exclude areas with substantially less exposure time, and therefore lower sensitivity. The exact value of the thresholding we use varies for each image but is selected to exclude the image edges. Problems with detections at the image edges are often noticeable as sharp rises or dips in the final radial density profiles of sources and occur either at image gaps, for cases where multiple images are used to cover the field around a galaxy, or at the largest radii. We guide our selection of the threshold value by adopting the smallest threshold value that eliminates such features. 

\begin{figure}[t]
\epsscale{1.0}
\plotone{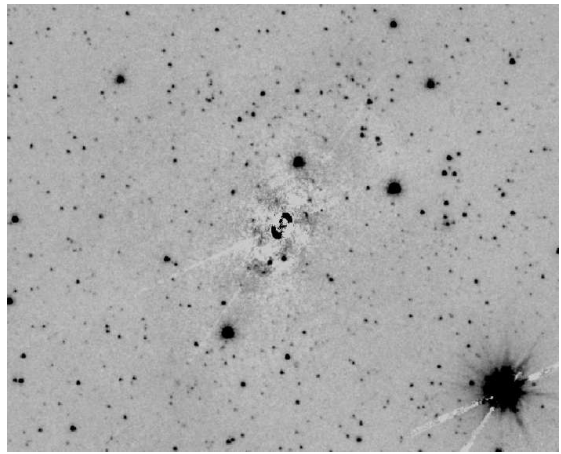}
\caption{Demonstration of the model subtraction near the core of NGC 1553.  The quality of the model subtraction is highly variable within a few kpc of the galaxy center.}
\label{fig:modelzoom}
\end{figure}

The basic pre-processing steps include sky subtraction plus modeling and removal of the primary galaxy.
We calculate the background sky value by evaluating the median within either the upper or lower quarter of the image, depending on whether the primary galaxy lies in one or the other of these two regions, of unmasked pixels. We subtract this median sky value from the entire image. We then use the IRAF$^{}$ task ELLIPSE to measure the properties of the central galaxy, create an image of that model using BMODEL and then, by subtraction, obtain an image that is as nearly free of the primary galaxy as possible (see Figures \ref{fig:modelsub} and \ref{fig:modelzoom}). We choose to use the ELLIPSE task, with free ellipse parameters, rather than the more detailed model fitting of S$^4$G galaxies \citep{salo} using GALFIT \citep{peng} or BUDDA \citep{desouza} because the ELLIPSE fitter proved to be superior at removing the smooth galaxy light. The GALFIT or BUDDA fits provide a more physically motivated approach and estimates of the parameters of physically distinct components in these galaxies (e.g. bulge and disk), and thereby enable one to address a wide range of questions, but for our purpose, which is to remove as much of the galaxy light as possible, the freedom of the model-unconstrained ellipse fitting results in smaller residuals in the subtracted image. 

Examples of the galaxy subtraction both on the scale of the full image and expanded about the target galaxy are shown in Figures \ref{fig:modelsub} and \ref{fig:modelzoom}, respectively, for NGC 1553. This galaxy is among the larger (in angular extent) among our sample. The multipolar residual pattern seen in Figure \ref{fig:modelzoom} is typical, but stronger in galaxies with disks, and usually spans $\sim 1$ kpc in projected radius from the center. Within the central region there are limited areas where individual sources can be identified quite close to the center, as well as others where the modeling errors severely increase the local noise. Beyond this inner region, the modeling clearly works quite well. Our procedure for measuring completeness will account for this variation in detection sensitivity.

\begin{figure}[t]
\plotone{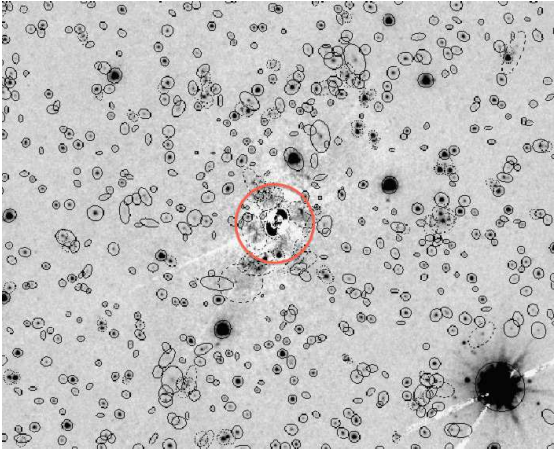}
\epsscale{1.0}
\caption{Demonstration of SExtractor object detection near the core of NGC 1553. We detect point sources well within galaxy. Some spurious sources are also created by the poorer 
model subtraction at projected separations  $<$ 3 kpc  a radius of 2 kpc is represented by the large central circle. The spurious sources are subsequently rejected either because they are 
identified as extended sources or not matched in the 4.5$\mu$m photometry (see text for details). 
Beyond the innermost central region, the source detection is of similar quality at all radii.}
\label{fig:objcheck}
\end{figure}

Once the residual images are available, we run SExtractor \citep{bertin} to identify point sources, eventually using the stellarity index to reject extended sources. When running SExtractor, one defines the criteria for an acceptable source using a specified minimum required number of pixels detected above a specified threshold, where that threshold is defined in terms of $n\sigma$ above background, where $\sigma$ is the background rms. We found that for a number of our images, SExtractor was miscalculating $\sigma$ because of the odd shape of the image footprints within the overall rectangular ``image". We therefore define a flux level that we consider significant and then evaluate the corresponding $\sigma$ threshold to reach that flux (over a minimum of two pixels). We evaluate this threshold interactively to reach the level of detection shown in Figure \ref{fig:objcheck}, guided by the criterion that detections be visually robust sources. We eventually remove most objects near this detection threshold from our catalogs when we set a uniform absolute magnitude limit, but that step is discussed below.  Catalogued objects that are clearly spurious, which tend only to be found near the galaxy center, are generally extended and so removed on the basis of that criterion, but we also use the 4.5$\mu$m images to help remove those as described below.

To remove extended sources, we opt to remove only unambiguously extended sources and accept all detections with stellarity $>$ 0.1. There are various ways to select between unresolved and extended sources, guided by the concentration or surface brightness of the image. However, because of the limitation of these data (2$^{\prime\prime}$ FWHM PSF), we are not in a position to resolve between faint background galaxies and point sources and therefore use only a basic measure of morphology and err on the side of inclusion.

Images of these galaxies at 4.5$\mu$m are also part of S$^4$G, and therefore exist for all of our targets. However, due to how observations were structured, these images are typically rotated 180$^\circ$ on the sky. This rotation means that while there is overlap between the 3.6 and 4.5$\mu$m imaging at the location of the target galaxy, the outlying regions, which are critical for determining the background source density, do not have overlapping coverage. For this reason, we cannot use the 3.6-4.5 color as a selection criteria  \citep[although it is known to be a weak diagnostic for star clusters in any case;][]{spitler08a}. However, in the region near the central galaxy, where our model subtraction is more uncertain, we use this additional band to help us discriminate against spurious sources. We run SExtractor in 2-image mode, using the 3.6$\mu$m image as the reference to photometer sources in the 4.5$\mu$m image that lie within 100 pixels of the central galaxy. This is the critical region where the model subtraction is most variable. Because the photometry can have large uncertainties in this region, and because we do not want to select against real source (even if they are not clusters - because we cannot enforce the same selection throughout the full 3.6$\mu$m image), we accept sources within a large color range $|3.6-4.5| < 2.5$ mag. We also apply the same procedure using a 5 mag color cut, but see no appreciable difference in the results (see below for more details).

\begin{figure}[th]
\plottwo{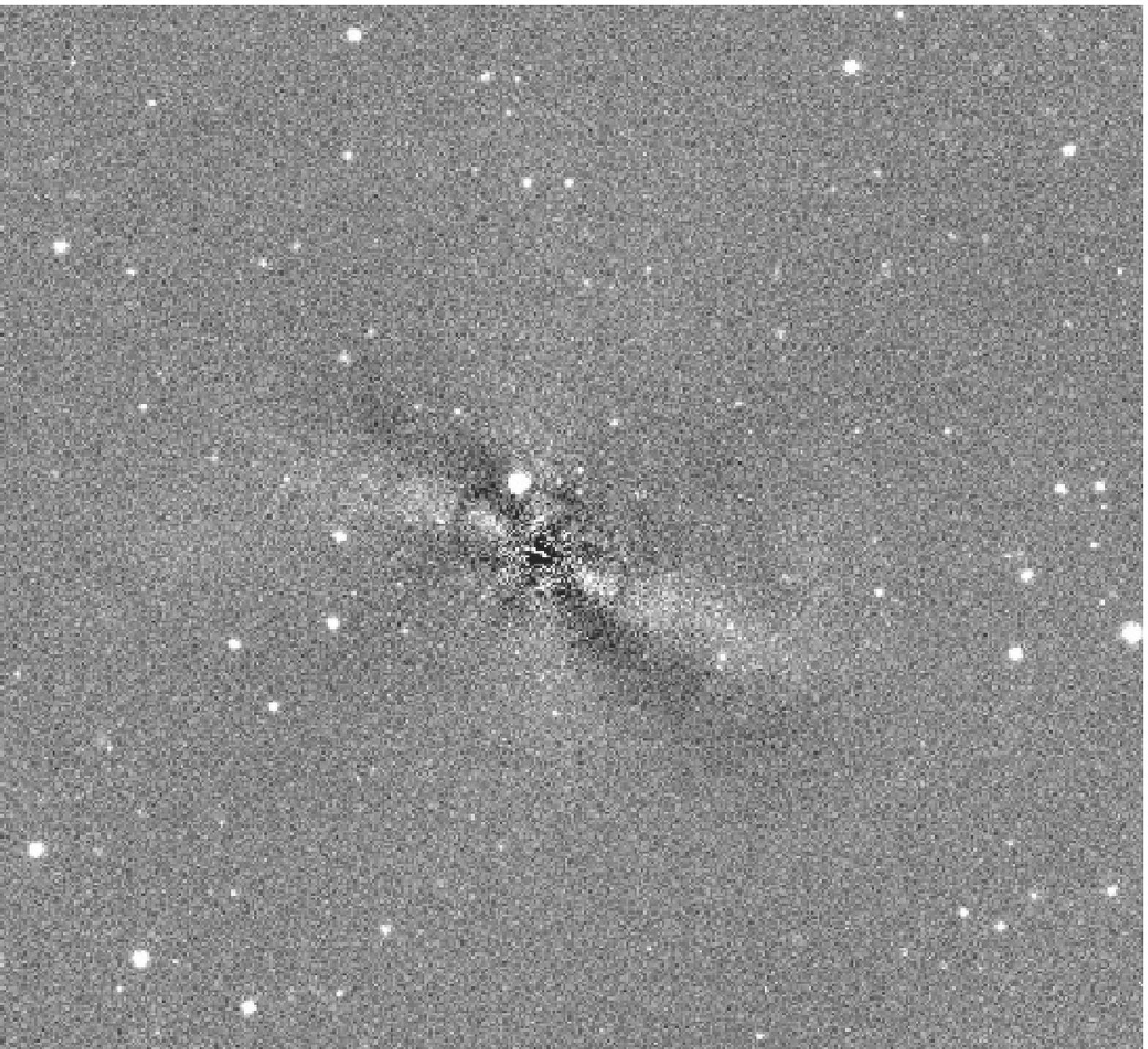}{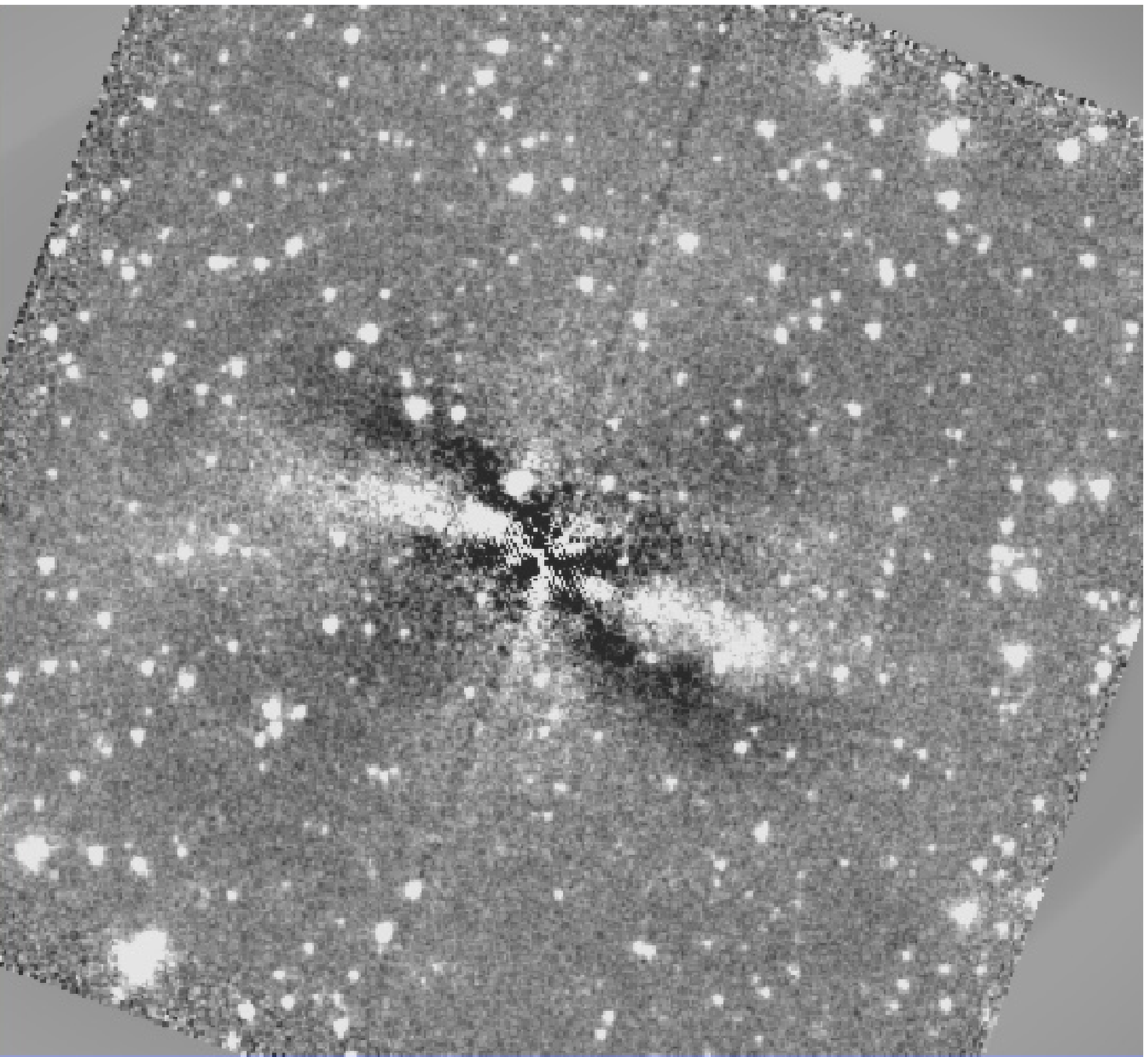}
\caption{Comparison of SDSS (left panel) and S$^4$G (right panel) residual (model subtracted) images for NGC 584. The point source density is manifestly much higher in the S$^4$G images. Here we select a different galaxy than in the previous plots both because an SDSS image of NGC 1553 does not exist and to provide a second, different example of a residual image. Sources can be detected close to the nucleus along the minor axis, but are lost in the residuals along the major axis. The completeness simulations will correct for the variable sensitivity using our simulations, under the assumption that the clusters are spherically distributed. A second image, offset to the northwest, exists but is not shown here and provides coverage to much larger radii. North is up and the angular size of the image portion shown is roughly 5 arcmin.}
\label{fig:compimage}
\end{figure}

\begin{figure}[th]
\epsscale{1.0}
\plotone{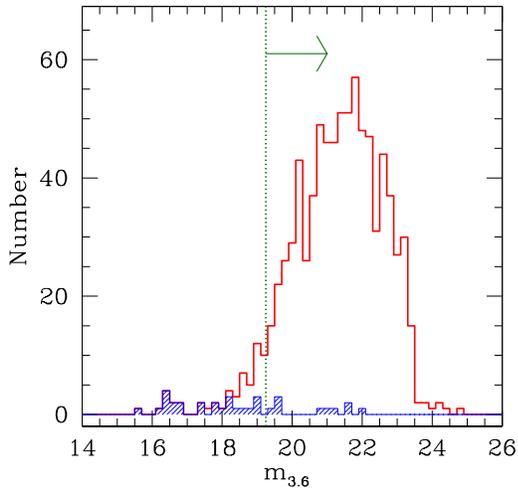}
\caption{Histogram of the apparent magnitude distribution for the combined source distribution of NGC 584 and NGC 1047. We used two galaxies to increase the statistics and ensure the result was not related to the selection of an inappropriate galaxy. Otherwise, these are just the first two NGC galaxies from our sample in a numerically sorted list for which SDSS data are available. The red unshaded histogram shows the S$^4$G source distribution, while the blue shaded histogram shows the matched SDSS sources. The vertical dotted line and arrow indicate the apparent magnitudes at which we expect to find globular clusters, set at a value that is 2$\sigma$ brighter than the peak magnitude of the globular cluster luminosity function (LF) and $\sigma$ is the dispersion adopted for the Gaussian LF (see \S 2.2 for discussion of the cluster LF). For the full range of distance moduli in our sample, the vertical line will range $\pm$ 1 mag from the plotted position. The SDSS images are grossly incomplete at the relevant magnitudes.}
\label{fig:comphist}
\end{figure}

Optical photometry would aid us either in confirming sources as real, or better still, in differentiating star clusters from other sources.
For example, optical-IR colors have been used to discriminate clusters from contaminants and to make further measurement of cluster metallicities \citep{kissler-patig,spitler08a}, and so combining our images with SDSS images, which are available for a large portion of the sky, is a natural avenue forward. However, the SDSS images turn out to be insufficiently deep. In Figure \ref{fig:compimage} we show the galaxy-subtracted residual images for NGC 584. It is evident from this comparison that the 3.6$\mu$m images go much deeper than the optical images. Of course, this difference would not be relevant if either the SDSS image was sufficiently deep to detect the clusters, or neither image was sufficiently deep. However, we show in Figure \ref{fig:comphist} that the apparent magnitude distribution of the SDSS data is grossly incomplete at the relevant magnitudes and that the S$^4$G sample is well matched to reach the top few magnitudes of the globular cluster luminosity function. We conclude that we are unable to use the SDSS data to help in our selection and that completeness corrections will not be extreme for the S$^4$G cluster counts. A deep set of optical images that cover the footprints of the S$^4$G galaxies will provide value in revisiting this question and enable one to examine the subpopulations of clusters in these galaxies. 

Given the lack of additional data to aid in selecting clusters, we implement a basic 3.6$\mu$m magnitude cut to exclude sources that are clearly too bright to be clusters at the distance of the target galaxy or that are sufficiently faint that we are beginning to reach within the highly incomplete range of our data. Guided by these limits, we constrain ourselves to sources with $-11 < M_{3.6} < -8$, the lower limit arising from defining a cut that is above the cluster detection limit for all of our galaxies.

Constructing the point source catalog is only the first necessary part of our measurement. As a function of both magnitude and location within the image, the completeness of our source counts will vary. To determine completeness, we add artificial point sources over a range of magnitudes that is greater than that defined for the candidate clusters (because measurement uncertainties could move objects within our magnitude limits).  
We cannot add too many sources without affecting the statistical properties of the image (where artificial sources start to overlap other artificial sources), so we rerun the process 50 times. To obtain better sampling of the regions near the galaxy, we distribute the artificial sources using a Hubble profile with a core radius of 100 pixels. We do assume the clusters are spherically distributed about the galaxy, but this is again an area where guidance from existing work is lacking. Some studies even suggest that the distribution varies for the different cluster populations \citep{wang}, complicating this issue even further. Within apparent magnitude, we distribute the sources uniformly.
The images with the artificial sources are then processed and analyzed in the same manner as the original ones.
 For each radial bin, we bin  by 25 pixels, we evaluate both the number of real sources and for each of the 50 realizations the fraction of artificial sources that is recovered (for sources within the absolute magnitude range of $-$11 to $-$8, assuming all sources are at the distance of the primary galaxy). This procedure results in enough sources, even at the smallest radii, that the uncertainties in our incompleteness corrections are subdominant over counting statistics. We then correct the observed number counts and, when converting these numbers to a surface density, we correct for the missed pixels beyond the image boundary by using only the area of the annulus within the image. The correction for masked pixels within the image boundary comes from our artificial source recovery fractions.

\subsection{Measuring the Cluster Populations}

Using the radially binned, completeness corrected surface source density values, we now estimate the parameters of a power-law profile description of the cluster distribution. The data are insufficient to allow for the fitting of the power law and background simultaneously. We adopt two approaches. First, we measure the background projected density level using the surface density values at radii $ > 30$ kpc and then fit a power law plus background model, with a free power-law slope and normalization, for radii between 1 and 15 kpc. Second, we  fix the power-law slope (the specific value chosen will be discussed below) and vary the power-law normalization and background level, still fitting the model for radii between 1 and 15 kpc. 
The fits are done by minimizing $\chi^2$. The choice of one model over another hinges conceptually on whether the background at large radii is a sufficiently accurate description of the background at smaller radii. We will select our preferred approach by comparing our results to previously published results for the limited subsample of galaxies where such measurements are available.

The power-law description for the radial distribution of globular clusters is historical \citep[see][]{brodie}, but fails in detail to describe known systems when very large samples of clusters define the radial profile precisely \citep[for examples of truncations, see][]{rhode01,dirsch}. Nevertheless, for the quality level of our data and in the interest of homogeneity in analysis, the power-law is adequate and preferred. As shown in Figures \ref{fig:profiles} and \ref{fig:profiles2}, the power-law model does a satisfactory job of fitting almost all of the galaxies.

The resulting values for the background (logarithm of the counts per bin; Back.), power law slope ($b$), and normalization ($a$) for the models where the background is set and power law index is free are given in Table \ref{tab:results}, along with the distance modulus (DM), Hubble T-Type (T), the 3.6 and 4.5$\mu$m apparent magnitudes, and the quality flag (Q). The integrated number of clusters out to a radius of 50 kpc (N$_{50}$), and the specific frequency relative to the galaxy's stellar mass (T$_{\rm N}$) corresponding to N$_{50}$ in units of number per 10$^9$ M$_\odot$ (as introduced by \cite{zepf}) is given for the models where the power-law index is fixed. The data and fits, shown including the background, are shown in Figures \ref{fig:profiles} and \ref{fig:profiles2} for all 97 galaxies, for fixed background and fixed power law index, respectively. 

\begin{figure*}
\epsscale{1}
\plotone{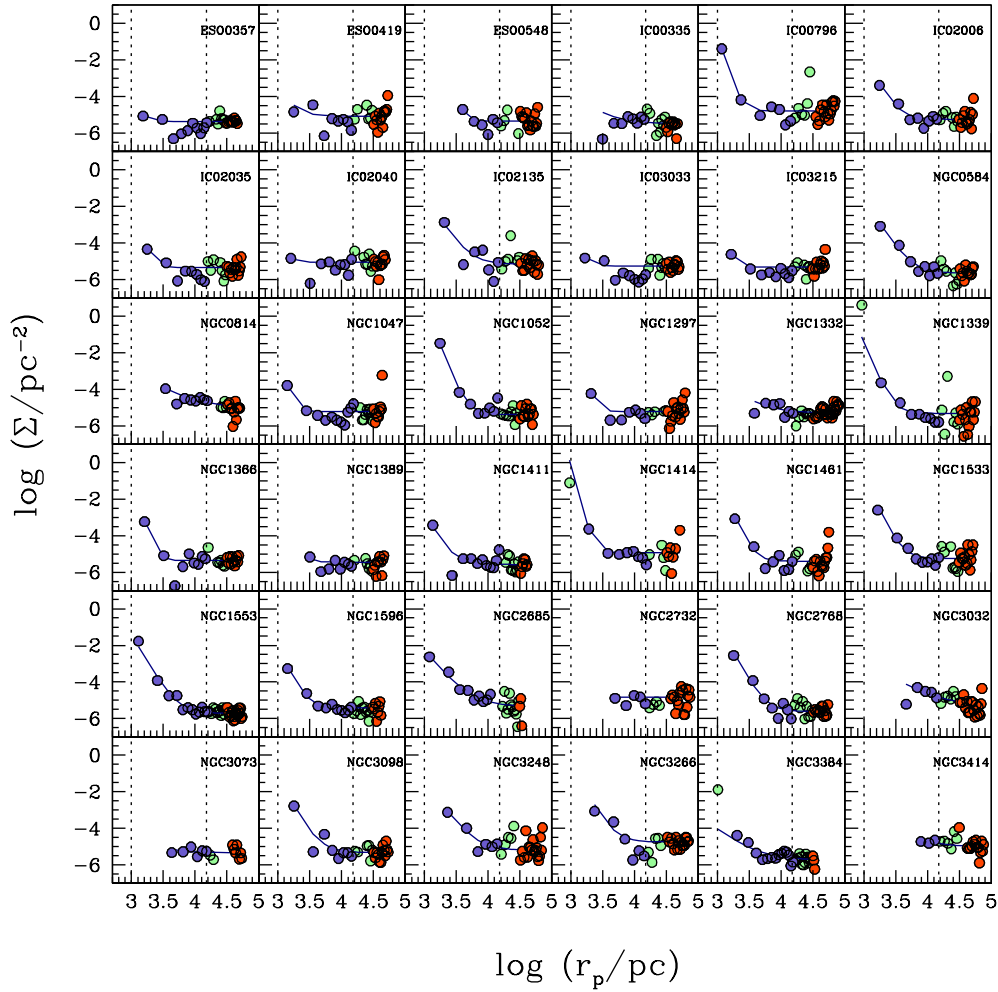}
\caption{Surface number density radial profiles of candidate globular cluster populations. Each panel contains the data for one galaxy.  The two vertical dotted lines denote the radial range over which the power-law model is fit (1 to 15 kpc). Data within that range are plotted as blue circles. The red circles denote the data used to determine the background source level and includes all data beyond 30 kpc. Data that are neither in the fitting range or background range are plotted as light green. The solid line shows the best fit model plus background over the radial range for which data exist. These plots represent the results of models where we set the background to be the measured surface density beyond 30 kpc as described in the text and leave the power law slope and normalization as free parameters.}
\label{fig:profiles}
\end{figure*} 

\setcounter{figure}{5}
\begin{figure*}
\plotone{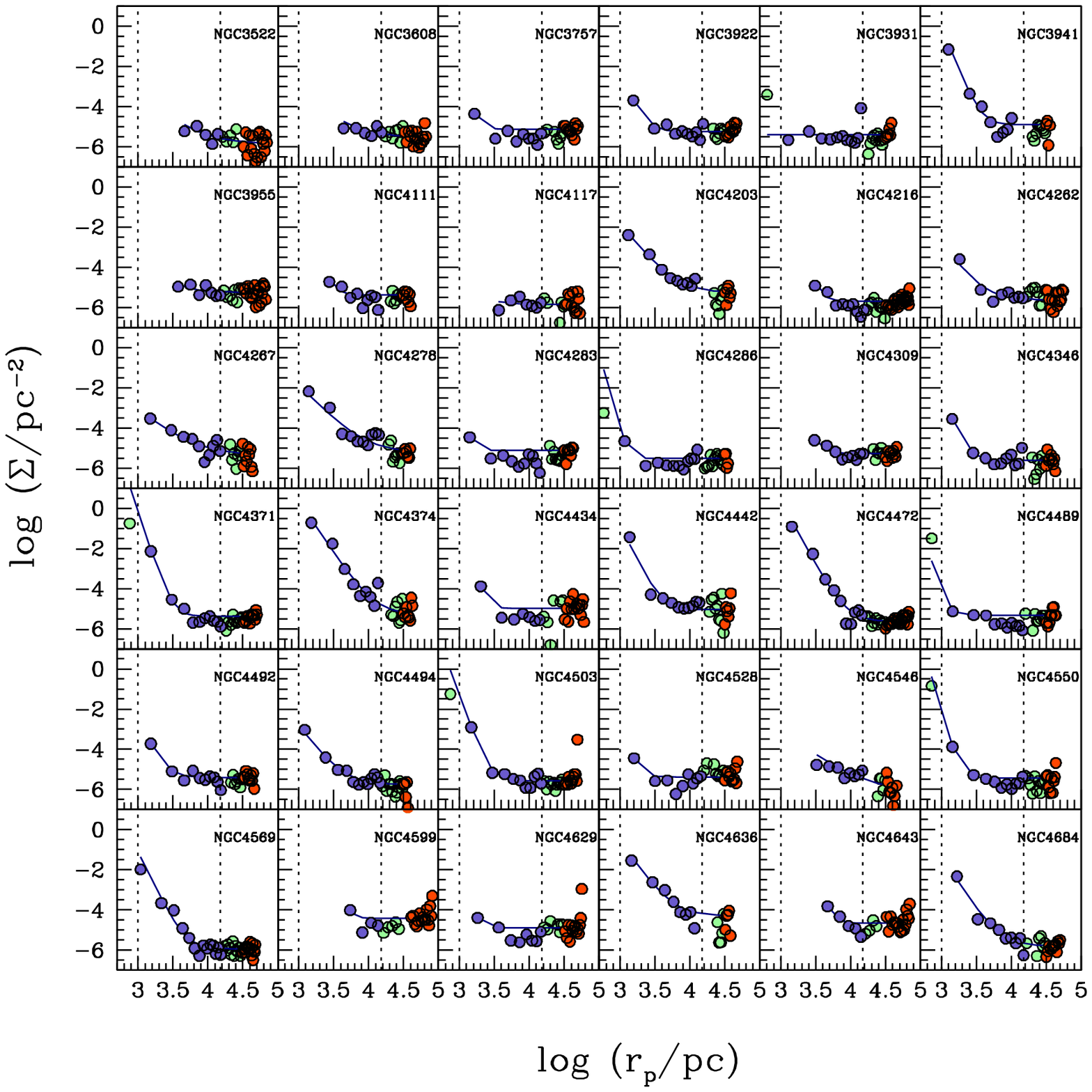}
\caption{cont.}
\end{figure*} 

\setcounter{figure}{5}
\begin{figure*}
\plotone{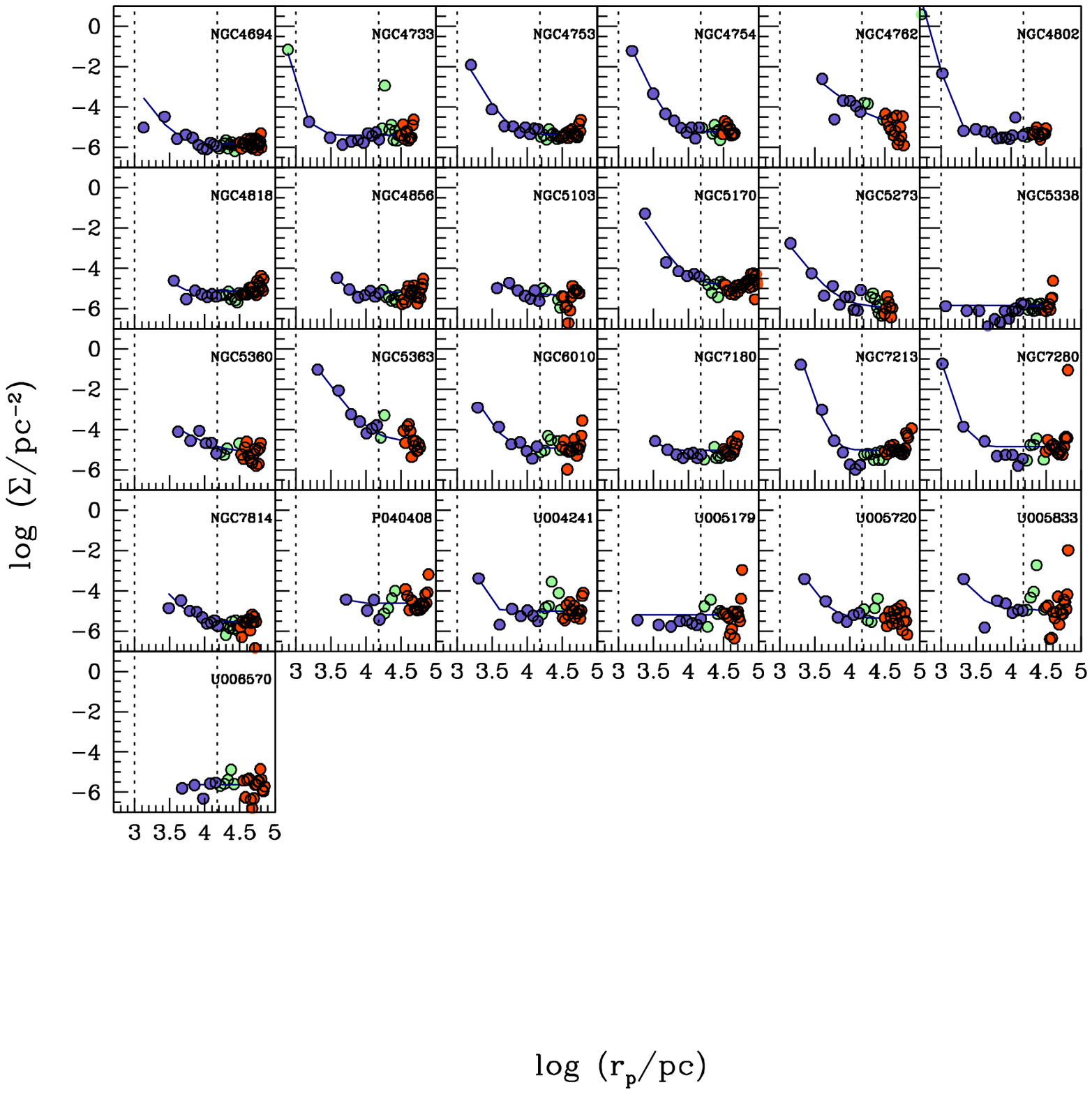}
\caption{cont.}
\end{figure*} 

\begin{figure*}
\plotone{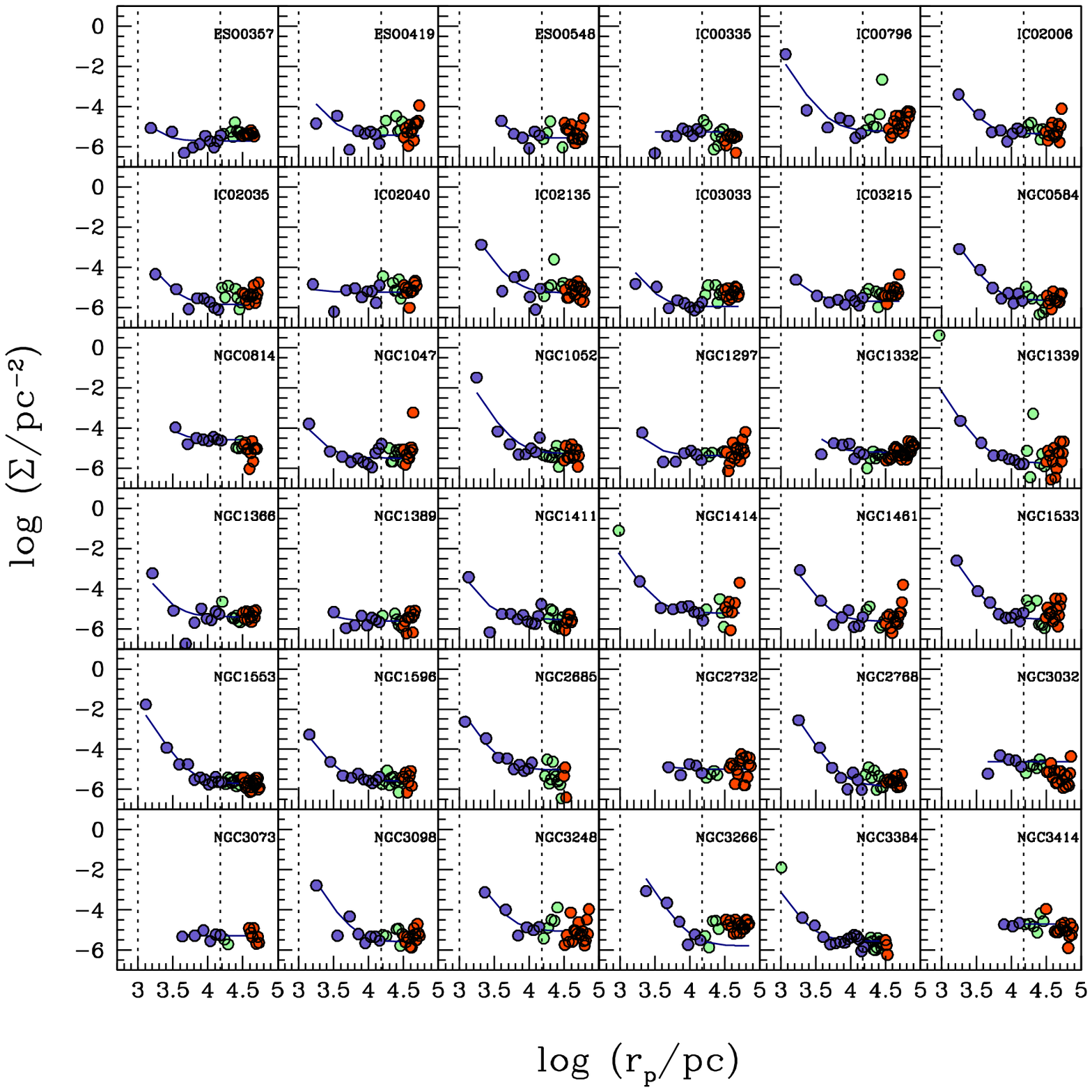}
\caption{Same as in Figure \ref{fig:profiles} except these plots represent the results of models where we fix the power law slope (see text) and leave the power law normalization and background as free parameters.}
\label{fig:profiles2}
\end{figure*} 

\setcounter{figure}{6}
\begin{figure*}
\plotone{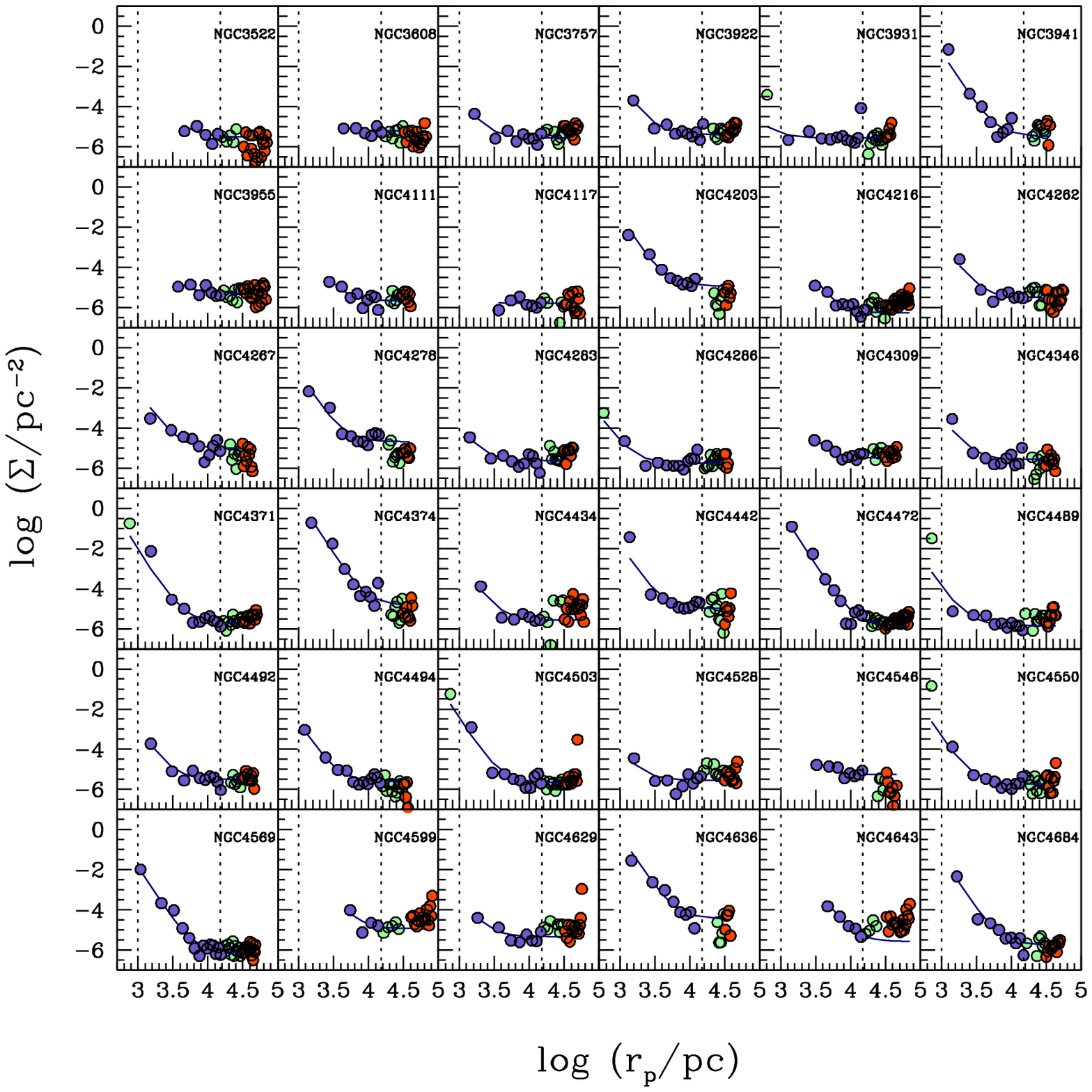}
\caption{cont.}
\end{figure*} 

\setcounter{figure}{6}
\begin{figure*}
\plotone{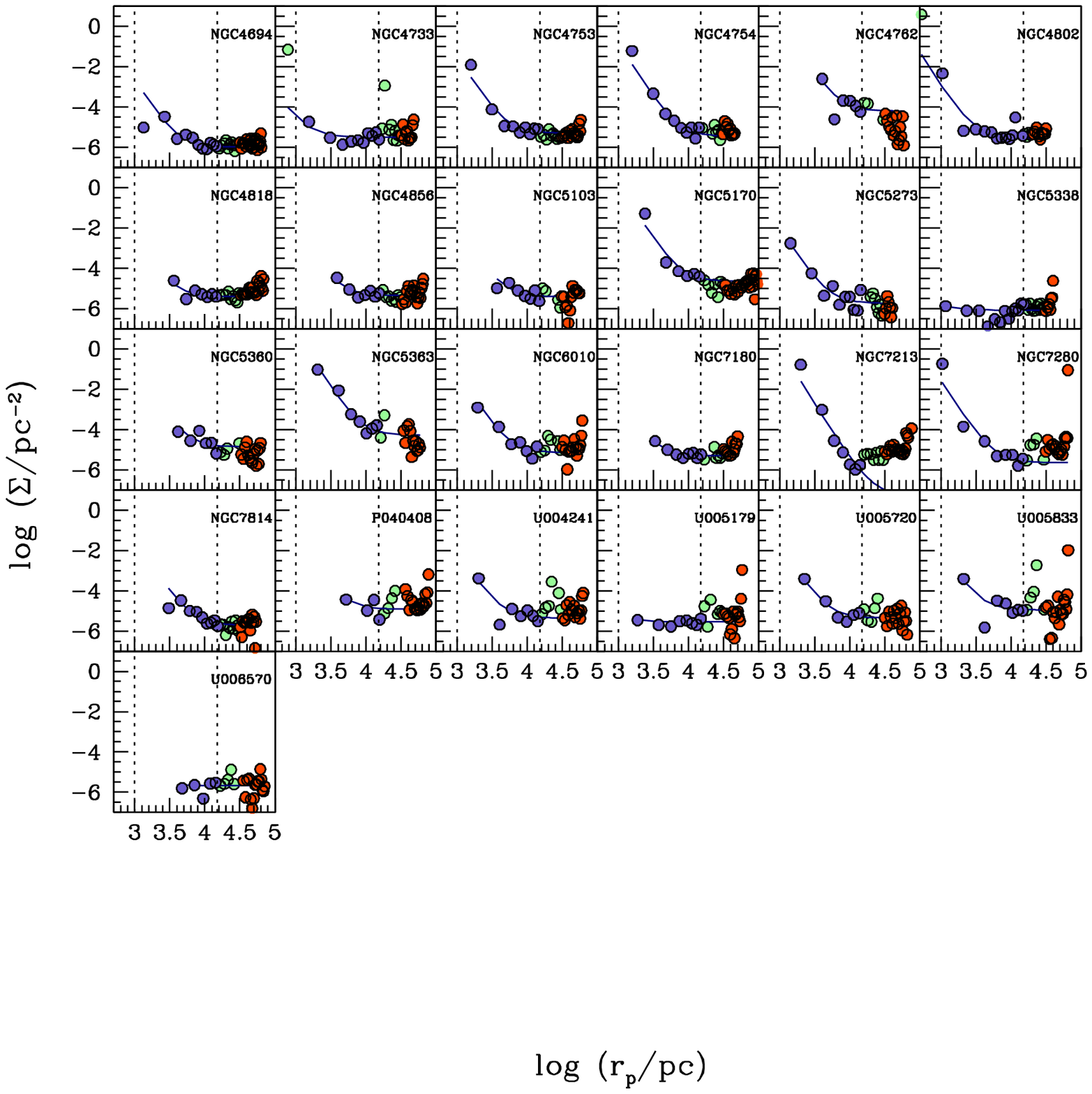}
\caption{cont.}
\end{figure*} 

Uncertainties in T$_{\rm N}$ are based on Poisson statistics in the individual radial bins, propagated through the fitting using $\Delta \chi^2$. In cases where the model fit is statistically acceptable we adopt the uncertainties corresponding to $1 \sigma$ in the model parameters. In cases where the model is statistically unacceptable, for the adopted Poisson uncertainties in the individual bins, we calculate the $\chi^2$ for which we would have only a 50\% chance of rejecting the model fit and rescale all of the uncertainties by the required amount to reach this 
$\chi^2$. The justification for this rescaling is that there are systematic uncertainties in the background that are not captured in the statistical uncertainties. We then evaluate the model parameter uncertainties by defining the $1 \sigma$ ranges using the larger bin uncertainties. We will validate these uncertainty estimates by comparing to literature values of N$_{\rm CL}$ in \S \ref{sec:literature}. 

Evaluating models on the basis of $\chi^2$ only judges models where data exist, but the data may not exist over an interesting range of parameter space. 
Another way of judging our profiles is based on how well the data that constrains them span the key radial range of 1 to 15 kpc. In some cases, the data span only a minority of this range and so even if the models are statistically valid for these galaxies, they will have large associated uncertainties due to the lack of good constraints on the inner profile. To quantify this distinction in profile quality, we consider the fits to those galaxies for which the data do not reach interior to $\log r = 3.5$ (3.1 kpc), to be of lower quality. We designate these profile fits, and those of the few galaxies that show highly irregular profiles (such as NGC 4762), as $Q = 0$ galaxies. This quality index is included in Table \ref{tab:results}.

\begin{figure}[t]
\epsscale{1.0}
\plotone{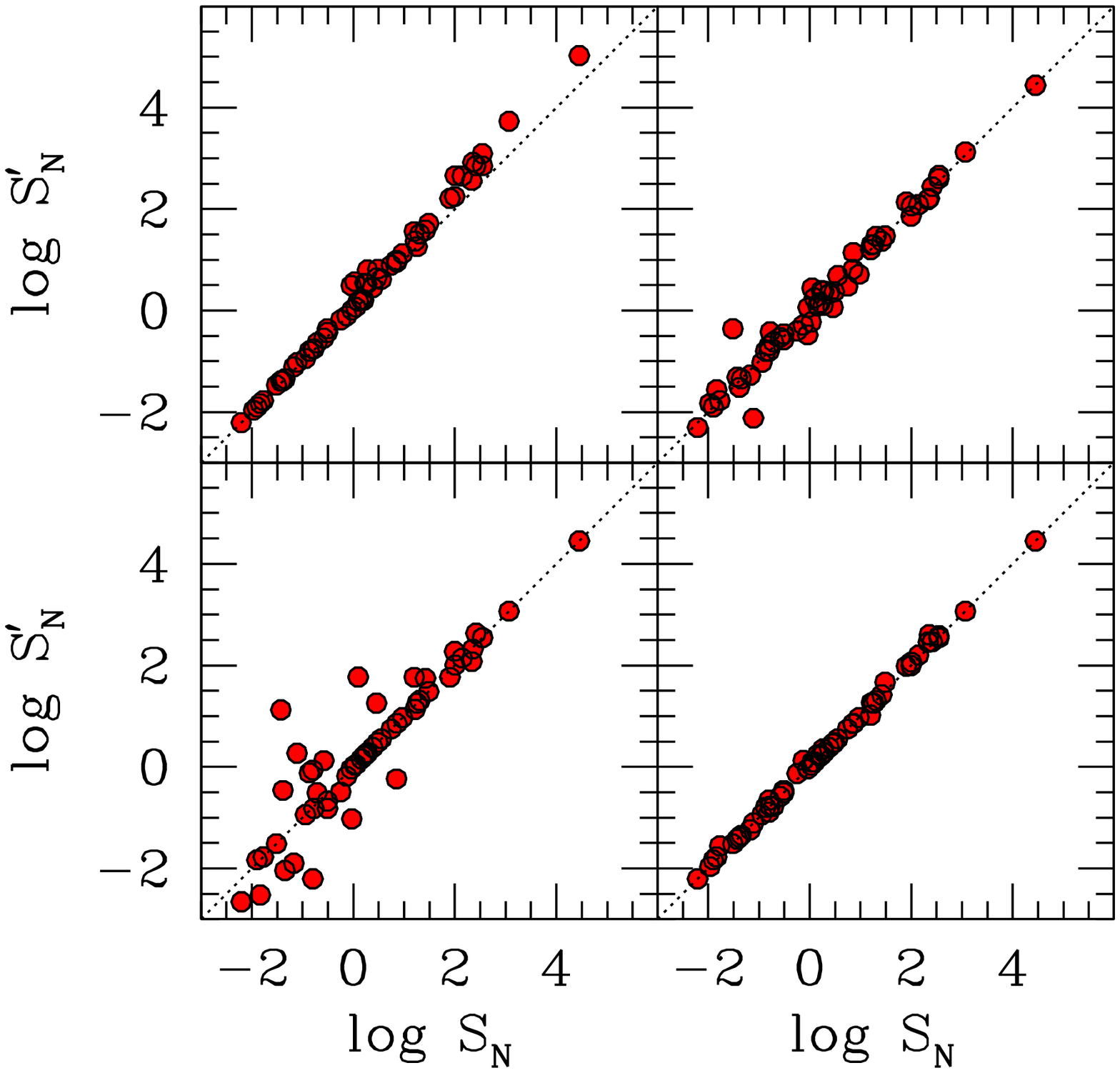}
\caption{Test of various aspects of our procedure. S$^\prime_{\rm N}$ denotes the ``new" version of the specific frequency and S$_{\rm N}$ is our standard version. In the upper left panel we consider how S$_{\rm N}$ changes if we integrate our fitted profiles to 100 kpc instead of the standard 50 (for models with free power law slope). In the upper right panel we calculate the new values by utilizing 100 simulations from which to derive our completion corrections rather than the standard 50. In the lower left panel we consider the effects of fitting to the data only between $2 \le r_p ({\rm kpc}) \le 15$ rather than $1 \le r_p \le 15$. In the lower right panel we consider the effect of increasing our color cut for spurious sources from 2.5 to 5 mag. None of these cause large changes to the bulk of the results, particularly at S$_{\rm N} > 1$, where the statistical errors are somewhat smaller.}
\label{fig:probcomp}
\end{figure}

To quantify the number of clusters in each galaxy we would in principle integrate our model profile to $r = \infty$. However, because we only empirically constrain the profile within 15 kpc, we are reticent to extrapolate the best fit profile far beyond this radius, particularly when we let the power law slope float.  Unfortunately, 15 kpc is clearly too small an outer radius to adopt if we intend to have a measurement of all of the clusters. In the Milky Way, one would be fairly close to the total number if one counted all the clusters within 50 kpc, even though there are a few clusters beyond 100 kpc \citep{harris79}.  Our aim is to obtain a measurement of the total number of clusters, not the number of clusters within a fixed physical radius because the latter will introduce a dependence of T$_{\rm N}$ on the physical scale of the cluster system. We show in the upper left panel of Figure \ref{fig:probcomp} that the numbers of clusters obtained by integrating to 50 and 100 kpc for models with floating power law slope differ modestly ($\sim$ 20\%), with only a slightly detectable systematic variation with the richness of the cluster system (models with fixed slope will have a fixed ratio between N$_{50}$ and N$_{100}$). We choose, therefore, to avoid the problem of extrapolating the fitted model to 100 kpc and treat the integral out to 50 kpc as the global number of clusters --- noting that this might be a slight systematic underestimation. We will return to this issue when we compare our results to those in the literature.

To obtain the number of clusters in each galaxy we use the integrated profile to $r = 50$ kpc and then correct that number for clusters outside of the magnitude range of our detected candidate clusters assuming a Gaussian luminosity function. Standard parameters for the peak and width of the luminosity function are M$_V \sim -7.4$ and $\sigma_V = 1.4$ for early types and $\sigma_V =$ 1.2 for later types \citep{brodie}. Adopting an estimate for the V$-$3.6 color of $\sim$2.4 \citep{barmby} results in an estimate of the location of the LF peak at M$_{3.6} = -9.8$. The dispersion of the cluster LF is not well measured at 3.6$\mu$m, but given the decreased sensitivity to age and metallicity differences at this wavelength, we adopt the lower range of $\sigma$ estimates in the $V$ band, $\sigma_{3.6} = 1.2$.  We adopt the same luminosity function for all galaxies. There is little variation in the globular cluster luminosity function with galaxy luminosity \citep{strader}. Variations of these parameters, within reason, tend to change N$_{\rm CL}$ by tens of percent, rather than by factors of a few, which is what we shall conclude is the actual uncertainty in our measurement (see \S \ref{sec:literature}).

The specific frequency can then be defined consistently with previous definitions as the number of clusters per parent galaxy luminosity, although here we would be using the 3.6$\mu$m luminosity rather than the historical B  or V luminosity. However, because we expect the ratio of clusters to total stellar mass to be the more physically interesting measurement, we go further along this path by using the {\sl Spitzer} magnitudes and their calibration to stellar mass \citep{eskew} to calculate a mass-dependent specific frequency. We produce a measurement of the ratio of the number of clusters to stellar mass rather than to the luminosity in one photometric band, normalized so that the values are in a similar numerical range (T$_{\rm N}$ is in units of clusters per $10^9 {\rm M}_\odot$ following the suggestion of \cite{zepf}).

The \cite{eskew} method for estimating stellar masses comes from a region by region comparison of reconstructed star formation histories in the Large Magellanic Cloud (LMC), based on analysis of optical color-magnitude diagrams \citep{hz},  and {\sl Spitzer} 3.6 and 4.5 $\mu$m photometry from \cite{meixner}. Differences in stellar populations among the regions provide an estimate for the robustness of the mass estimates against such variations, and result in an uncertainty estimate of 30\% (presumably lower in whole galaxies which should average over the more extreme star formation histories seen in localized regions of the LMC). The fitting formulae presented by \cite{eskew} have been confirmed, apart from  uncertainties related to differences in the adopted stellar initial mass function, by an independent analysis of SDSS spectroscopy \citep{cybulski} and a detailed analysis of S$^4$G  photometry \citep{querejeta}.

Before proceeding to discuss the results, we expand a bit on our set of choices. First, we have shown that the choice of integrating our model profile to 50 or 100 kpc produces little change (upper row, left panel). We opt to remain conservative in our extrapolation. Second, we show in Figure \ref{fig:probcomp} a comparison of S$_{\rm N}$ obtained for the standard parameter choices and that obtained using 100 rather than 50 simulations (upper row, right panel). There is no systematic difference between the two values and only some slight variance at extremely low S$_{\rm N}$ ($< 1$), where statistical uncertainties dominate. We will continue with the results drawn from 50 random simulations for the completeness corrections. Third, we show in the bottom row left panel a comparison of our standard S$_{\rm N}$ measurement with that obtained using an inner cutoff of 2 kpc in our model fitting. This panel shows the largest scatter among the four panels, but the scatter is mostly again for S$_{\rm N} < 1$, where the statistical error bars dominate. Given that for most systems the smaller cutoff radius results in no significant change in our S$_{\rm N}$ measurement and that the extra data at small radii can help reduce fitting uncertainties, we proceed with an inner cutoff of 1 kpc in our fitting. Finally, we explore changing the color cutoff  used to remove spurious sources from 2.5 to 5 magnitudes in the lower row, right panel. Here we find almost no detectable difference between the resulting S$_{\rm N}$ measurements, indicating that our cutoff of 2.5 mag is not resulting in the rejection of real sources.

\subsection{Comparison to Literature and the Validation of Our Adopted Methodology}
\label{sec:literature}

To close this section, we compare, where we can, our estimates of N$_{\rm CL}$ to those in the literature. This comparison is not straightforward. Serious differences can arise in the results from the range of completeness corrections, both in terms of detection of sources in an image and then in terms of correcting for the entire luminosity function of clusters. Differences in the adopted distance can further complicate matters. It is often the case that the completeness corrections are not particularly explicit and certainly differ at least in detail (such as in the peak and width of the Gaussian luminosity function) with ours. Furthermore, some of the best data with which to identify clusters, from the {\sl HST}, suffer from the small field of view and therefore spatial completeness corrections must be made. In certain studies \citep{rhode04,spitler08b} ground-based data is combined with {\sl HST} to provide both the improved resolution in the core of the galaxies and the larger coverage beyond. For all of these reasons, it is difficult to construct a homogeneous comparison sample from the literature. 
These concerns echo those expressed previously regarding literature compilations in general, but we will show below that these concerns appear to be at a level of precision below that which we or the literature studies achieve.

\begin{figure}[tj]
\epsscale{1}
\plottwo{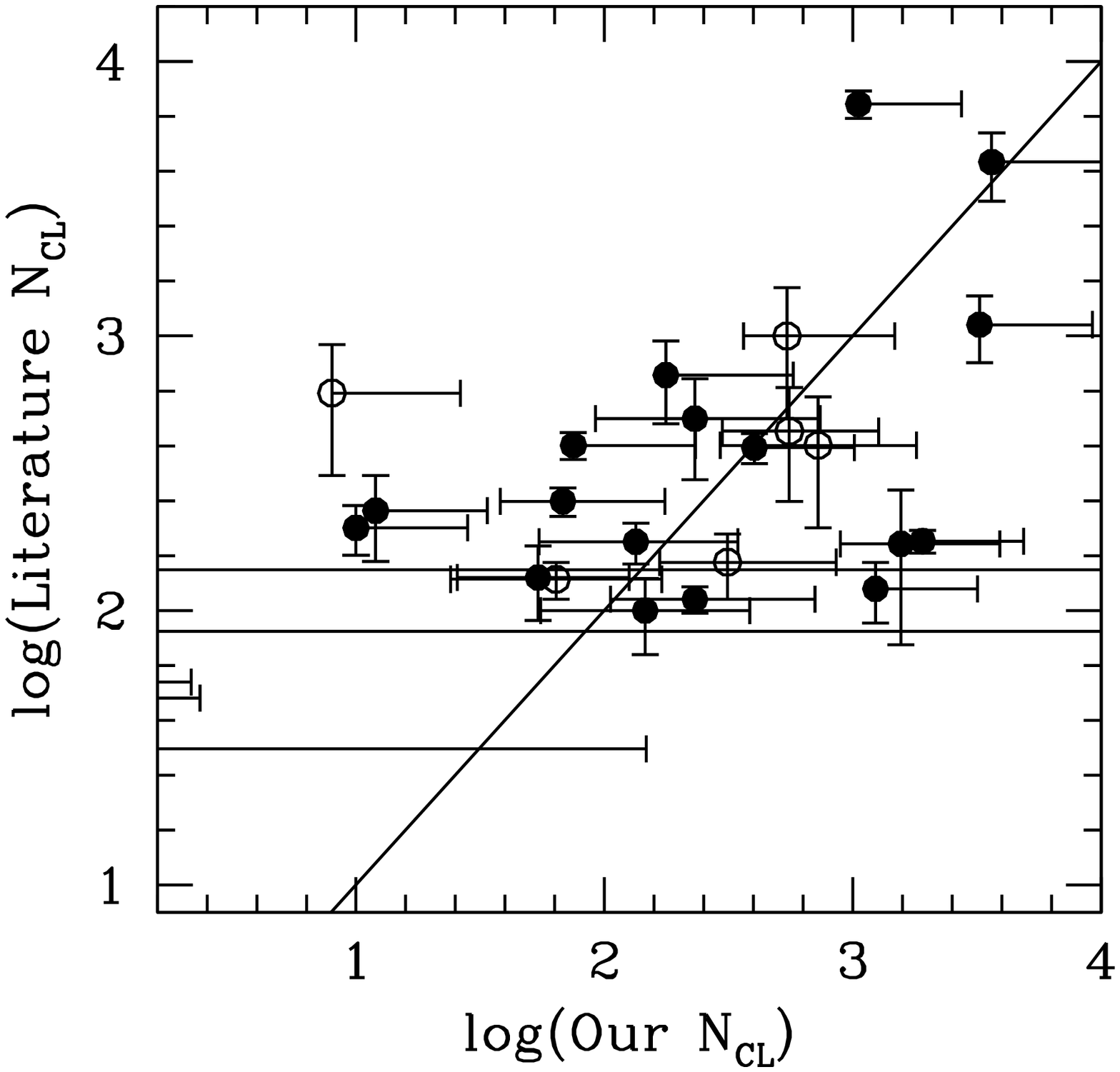}{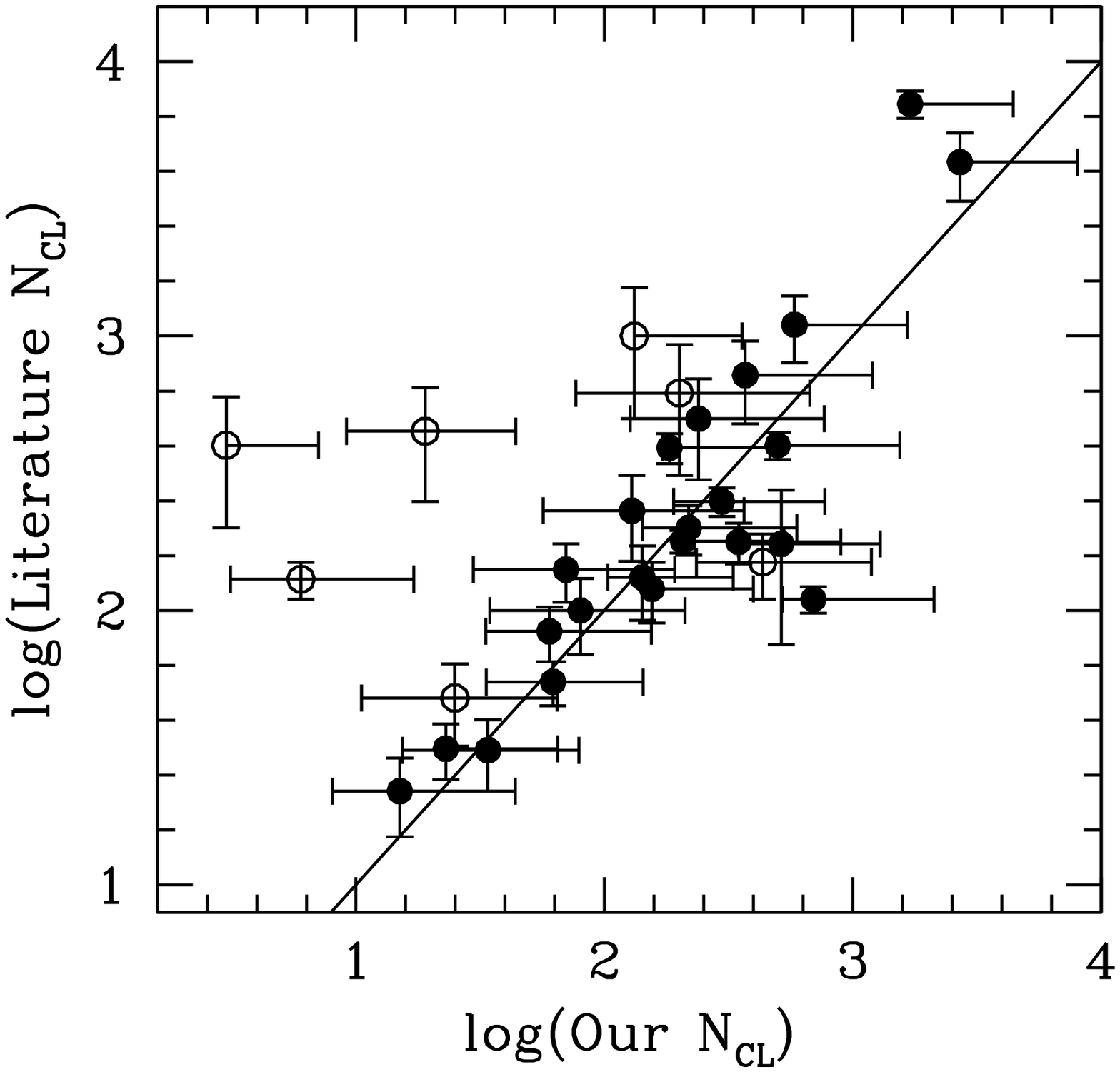}
\caption{Comparison of our values of N$_{\rm CL}$ and those in the literature.   The solid black line is the 1:1 relationship, not a fit to the data. Populations where we measure N$_{\rm CL} = 0$ have only their upper uncertainty limit shown. Uncertainties represent the internal, mostly statistical, uncertainties quoted by the literature studies and our own work. Left panel represents results from models where the background is held fixed to the value measured at large radius. Right panel represents results from models where the power law slope is held fixed to the preferred value of $-2.4$ (see \S\ref{sec:literature}). Open circles represent measurements with a quality flag of 0 (poor) and filled circles represent measurements with a quality flag of 1 (good). }
\label{fig:litcomp}
\end{figure}	

We use the results in the compilation of literature values by \cite{harris13} for the comparison shown in Figure \ref{fig:litcomp}.
We include in the comparison galaxies to which we had assigned a quality flag of 0 in order to maximize the sample size. Figure \ref{fig:litcomp} consists of two panels where we compare the results of the approach in which we fix the background to the measurement obtained at large radii (left panel) and  where we fix the power law slope (right panel).  The line plotted in both panels is the 1:1 line, not a fit. It does a satisfactory job of describing the mean trend, although there is clearly significant scatter about the line in the left panel, beyond that described by the internal uncertainties.

We find that the fixed slope modeling has two significant advantages over the fixed background modeling. First, the asymmetric tail of points at low N$_{\rm CL}$ that is seen in the left panel of the Figure is absent in the right panel. The tail in the fixed background measurements is presumably due to systems that are poor in globular clusters but whose background level is slightly overestimated by the measurement at large radius. In such cases, we  conclude that there are no clusters, whereas the more precise literature samples are able to recover the actual, but small, number of clusters. Second, the scatter about the line is visibly reduced using the fixed slope approach. Again, we suspect that systematic errors in the background estimation are what is driving the majority of the increased scatter in the left panel, although systems with poorly constrained power law slopes contribute to scatter because of the necessary extrapolation from 15 kpc to 50 kpc. Therefore, both because of background systematic errors and our inability to constrain the power law slope beyond $\sim$ 15 kpc, we conclude that the fixed slope approach is the more robust. With the exception of a few outliers, which mostly have a quality flag value of 0, the internal uncertainty estimates do a credible job of representing the scatter between our measurements and those in the literature, when we adopt the fixed slope approach.

For the fixed slope approach, the scatter about the 1:1 line is 0.29 dex for the $Q=1$ sample.
If we attribute all of the scatter to our measurements, this scatter suggests that the uncertainty in our measurements is roughly a factor of two. Although this is obviously a large number in absolute terms, relative to the range in N$_{\rm CL}$ of several orders of magnitude the uncertainty is relatively modest. The mean offset from the 1:1 line is only $-0.046$ dex (we   underestimate the literature values by 10\%) confirms that the use of N$_{50}$ as a proxy for N$_{\rm CL}$ is valid and that our 50 kpc integration limit produces a systematic error that is significantly smaller than the random errors. This quantitative agreement with the literature values in the mean also supports our decision to not include the correction for the background bias described by \cite{harris86} due to ``contamination" of the background estimates by the cluster population itself. We empirically determined the correction to be small in our data relative to our uncertainties. Hereafter, we refer to N$_{50}$ as N$_{\rm CL}$.

The results described above are valid for any choice of power law exponent between the plausible range of $-2$ to $-2.5$ in projected surface density valid for normal ellipticals \citep{brodie}. However, we have gone beyond that in producing the results in Figure \ref{fig:litcomp} in that we have chosen the value of the fixed power law index to minimize the scatter in the right panel. We obtain the result by recalculating N$_{\rm CL}$ for different choices of index, stepping in units of 0.1, and evaluating the scatter produced in the analog of the right panel of Figure \ref{fig:litcomp}. We find that we prefer a power law slope of $-2.4$ and adopt this slope universally in this study. The 3-D density will have a power law dependence that is one unit steeper, so the integral easily converges. 

We adopt a constant power law index, despite previous findings that the index depends on M$_*$ \citep{brodie}. If the radial slope does depend on mass, then fixing the slope could introduce an artificial trend in T$_{\rm N}$ with M$_*$. The previous claims are that the slope steepens as one progresses to lower mass systems. Such an effect could result in us underestimating N$_{\rm CL}$ in high mass galaxies and overestimating in low mass ones. However, when we examine the correspondence between our data and the literature values, the best fit slope obtained using the OLS bisector method presented by \cite{isobe} to account for uncertainties in both axes is $1.14\pm0.24$, and so consistent with the 1:1: line. We also find no significant correlation between our measured power law slopes and M$_*$ for log(M$_*/{\rm M}_\odot) > 10$ (below that mass our fits with free power law index are unreliable).
These results do not demonstrate the absence of such a relationship, only that it is too weak to detect with a sample of this size and scatter in N$_{\rm CL}$ that is a factor of two.

Similarly, we use the consistency of the literature comparison with the 1:1 line to argue that other simplifications we have adopted, such as the constancy of the peak magnitude of the cluster luminosity function and the fixed integration to 50 kpc regardless of galaxy size or mass, are not affecting our measurements at a level that is noticeable given the uncertainties. Such a statement could simply indicate that we have inferior measurements, after all the inability to measure an effect is not a desirable attribute, but as shown in Figure \ref{fig:litcomp} our estimated uncertainties are comparable or only slightly larger in general than those claimed by other investigators. Furthermore, in defense of our approach we cite the homogeneity of the sample and our algorithmic approach, which are critical in comparisons across galaxy luminosity, type, and mass.

The reader may still wonder why our measurements appear to be robust to variations in the radial power law slope and the peak magnitude of the cluster luminosity function (LF). The former likely arises because of the robustness of the integral under the model fits. Even with the wrong power law slope, the fit is likely to represent the mean surface density cluster values reasonably well over the fitted radial range and when we integrate over radius to arrive at a total cluster population we are relatively insensitive to the slope of the fit, as long as we evaluate the integral over a similar radial. As for the peak magnitude, our images are complete to below the peak magnitude, even with the anticipated possibility of a varying peak magnitude, and so our completeness corrections are never going to be larger than a factor of two, and clearly often much better. Because the observational scatter is a factor of two, variations due to improper corrections related to a variable peak LF magnitude are not easily detected.

\section{Discussion}

It is evident from the surface density profiles (Figures \ref{fig:profiles} and \ref{fig:profiles2}) that for $\sim$ 50\% of the galaxies there exists a clustered source population, which consists presumably of globular clusters, that a power-law is an adequate description for the radial distribution of these sources given the current state of the observations for most galaxies, and that there is a range in the properties of that population (numbers and radial extent) among our sample. Our argument that these sources are globular clusters is circumstantial because we lack the resolution to confirm their nature. We base our conclusion on 1) they evidently cluster about the parent in those cases where an excess is seen, 2)  an excess is not found 
in every case, even when the data extend to small radii, demonstrating that the sources do not spuriously arise from residuals in the parent galaxy subtraction process, 3) the sources have absolute magnitudes consistent with those expected for globular clusters, 4) the radial distribution of sources, where that is well measured, lie in the range of $r^{-2}$ to $r^{-4}$, which is consistent with previous measurements \citep{ashman98,brodie}, 5) the numbers of excess sources, ranging from a few up to a thousand, are in the range expected based on previous cluster population studies \citep{harris,brodie}, and 6), in the few cases where we can compare our measurements to higher fidelity measurements in the literature, we reproduce prior results. From now on, we will refer to these excess sources as clusters, although it must be understood that these populations could be partially contaminated by other sources that also cluster about the parent galaxy. In general, one possible such source would be star forming knots \citep{thilker,gdp,zc,hf} and intermediate age versions of such structures,  although for our early-type galaxies those should not be a major source of contamination. Individual stars are generally insufficiently luminous to match our magnitude criteria $(-11 < {\rm M}_{3.6} < -8)$, although some especially luminous stars \citep{blum}, again not as likely in early-type galaxies, could fall in this range. 

\subsection{The Numbers of Clusters}

\begin{figure}
\epsscale{1.0}
\plotone{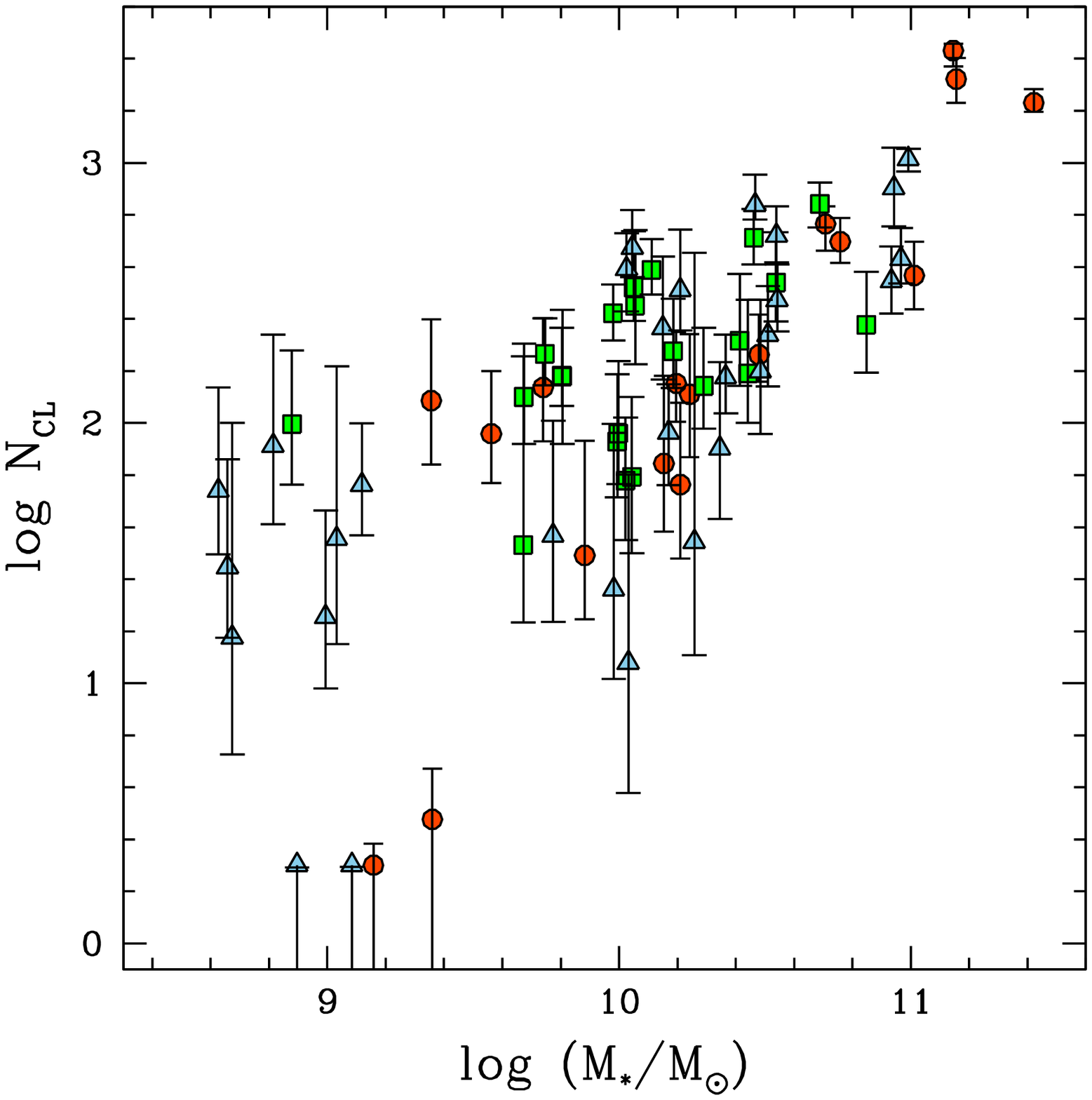}
\caption{Cluster population sizes (N$_{\rm CL}$) versus parent galaxy stellar mass. Red circles denote classic ellipticals (T $\le -4)$, green squares denote intermediate early types ($-4 < T < -2)$ and blue triangles denote later galaxies ($-2 \le T \le  1)$. Only galaxies with a quality flag of 1 and with photometry in both IR bands are included.}
\label{fig:numbers}
\end{figure}

\begin{figure}
\epsscale{1.0}
\plotone{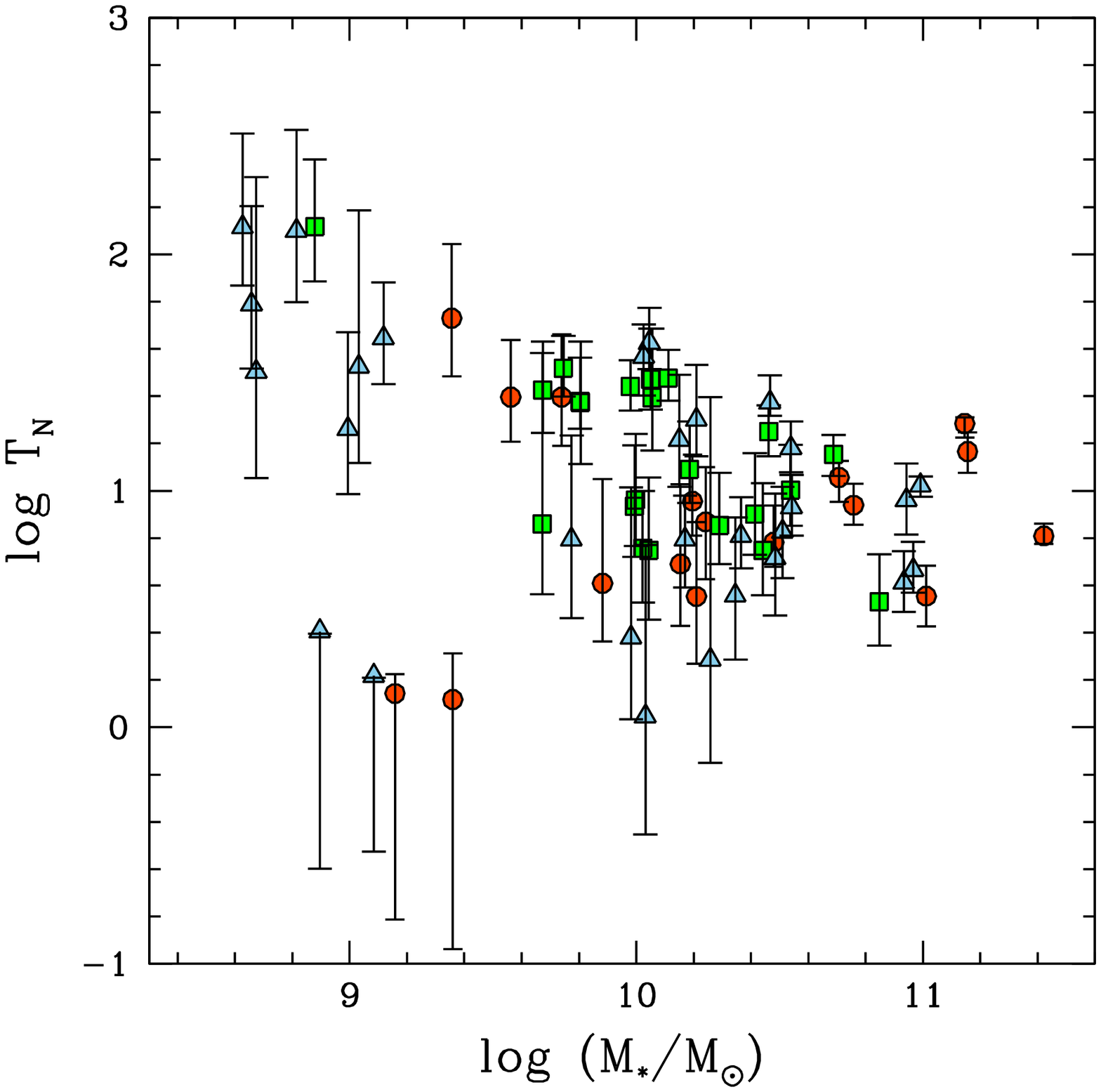}
\caption{Cluster population mass-normalized specific frequency (T$_{\rm N}$) versus parent galaxy stellar mass. Red circles denote classic ellipticals (T $\le -4)$, green squares denote intermediate early types ($-4 < T < -2)$ and blue triangles denote later galaxies ($-2 \le T \le  1)$. Only galaxies with a quality flag of 1 and with photometry in both IR bands are included.}
\label{fig:specfreq}
\end{figure}

We show the number of clusters, N$_{\rm CL}$, as a function of parent galaxy stellar mass, M$_*$ in Figure \ref{fig:numbers}.  Some basic qualitative results can be drawn quickly from the Figure: 1) there is a general increase in the N$_{\rm CL}$ that roughly follows the rise in stellar mass, 2) despite this mean trend, for M$_* < 10^{10}$ M$_\odot$, the variation in N$_{\rm CL}$ approaches a factor of 100 at a given M$_*$ and may reflect different galaxy characteristics, 3) for M$_* < 10^{10} {\rm M}_\odot$ the behavior of N$_{\rm CL}$ can be characterized as flat, and 4) that neither the trend at large M$_*$ nor the scatter at low M$_*$ correlate strongly with morphology (within this admittedly early-type sample).  We expand on each below.

A linear fit to the data using orthogonal regression \citep{isobe} results in a best fit slope of $0.94\pm 0.19$, supporting the qualitative suggestion that at least roughly N$_{\rm CL} \propto$ M$_*$. If we only use the data for log $({\rm M}_*/{\rm M}_\odot) > 10$, which avoids the more ambiguous lower mass galaxies, then the slope is $1.39 \pm 0.36$ ($1.24 \pm 0.19$ if using the bisector method). Slopes $>$ 1 suggest that T$_{\rm N}$ is increasing with M$_*$. This result is consistent with that of \cite{harris13} for this same mass range. The implication of such a result is that more massive galaxies are more capable of producing clusters. Such a result has repercussions for merger models in which clusters may form \citep{schweizer,ashman92} and for models in which merging is dissipationless. However, as we will discuss in the next section, such interpretations are premature until we understand the nature of the stellar initial mass function and its effects on our estimates of M$_*$. 

Among galaxies with log (M$_*/{\rm M}_\odot) < 10$, T$_{\rm N}$ exhibits large scatter, with values that can differ by as much as two orders of magnitude (Figure \ref{fig:specfreq}). These galaxies have limited populations of clusters, and therefore statistical errors in combination with subtle systematic errors may be responsible for the large scatter. The trend for at least some of these systems to have proportionally larger N$_{\rm CL}$ than expected for their mass, is clearly visible here and in the study of \cite{harris13}. Furthermore, our comparison to the literature values of N$_{\rm CL}$ suggests that we are obtaining reliable estimates of N$_{\rm CL}$, with plausible internal uncertainty estimates, even for galaxies with low N$_{\rm CL}$. Unfortunately, the object-by-object comparison to the literature is limited to a few galaxies with similarly low N$_{\rm CL}$ and so we may simply have been fortunate in those galaxies that overlapped existing studies.  However, other studies find significant scatter in specific frequency as well among low mass galaxies. For example, \cite{miller} find S$_{\rm N}$ values that range from 0 to 23 for a set of dwarf elliptical galaxies. The quantitative values are not directly comparable to our results because their normalization is done with respect to an optical luminosity, but the range in values is greater than an order of magnitude. Also, \cite{strader} suggest that there are two families of galaxies among the dwarf ellipticals, one with S$_N \sim 2$ and another with S$_N \sim 5 - 20$.  Again, the numbers are not directly comparable to ours because of the use of bluer bands and our conversion to stellar mass, but the suggestion of two families is consistent with the visual impression of our Figure \ref{fig:numbers}. The existence of two populations is not evident in \cite{harris13}.

We also find that there is marked change in the behavior of N$_{\rm CL}$ with M$_*$ below a galaxy stellar mass of $\sim 10^{10}$ M$_\odot$, consisted with previous studies such as \cite{georgiev} and \cite{harris13}. As we discussed when reviewing the increased scatter at low M$_*$, these systems are more susceptible to systematic errors. Nevetherless, there is little if any dependence of N$_{\rm CL}$ with stellar mass for log(M$_*/{\rm M}_\odot) < 10$, suggesting that some of the low stellar mass galaxies have the largest T$_{\rm N}$. It appears that large variations in T$_{\rm N}$, perhaps driven by environment or star formation historyf \cite[for examples of related phenomenon see][]{georgiev}, are present in galaxies with low stellar masses and that those differences average out as one examines galaxies with increasing M$_*$. Alternatively, these systems may also have lost much of their baryonic material (presumably not  through tidal stripping which would affect stars and clusters, but through stellar feedback which could have removed material destined to form stars but not clusters). As such, the range of T$_{\rm N}$ may help constrain models of stellar feedback \citep{dekel}. These speculations motivate more study of low mass galaxies  in understanding the drivers of cluster formation. Of course, increased verification of N$_{\rm CL}$ measurements in low M$_*$ galaxies would need to be part of such a program.

At the high mass end, we do not find the sharp rise in T$_{\rm N}$ found by \cite{peng08}. There are various possible reasons, some trivial some interesting, for this disagreement. First, we have only four galaxies in this mass range, so we may simply have been unfortunate (likewise \cite{peng08} have a small number in this mass range). Potentially more interesting is that the \cite{peng08} sample consists exclusively of Virgo galaxies and the increased number of clusters that they find around the massive galaxies
may be the result of stripping of clusters from lower mass galaxies in the cluster environment.

\begin{figure}[th]
\epsscale{1.0}
\plotone{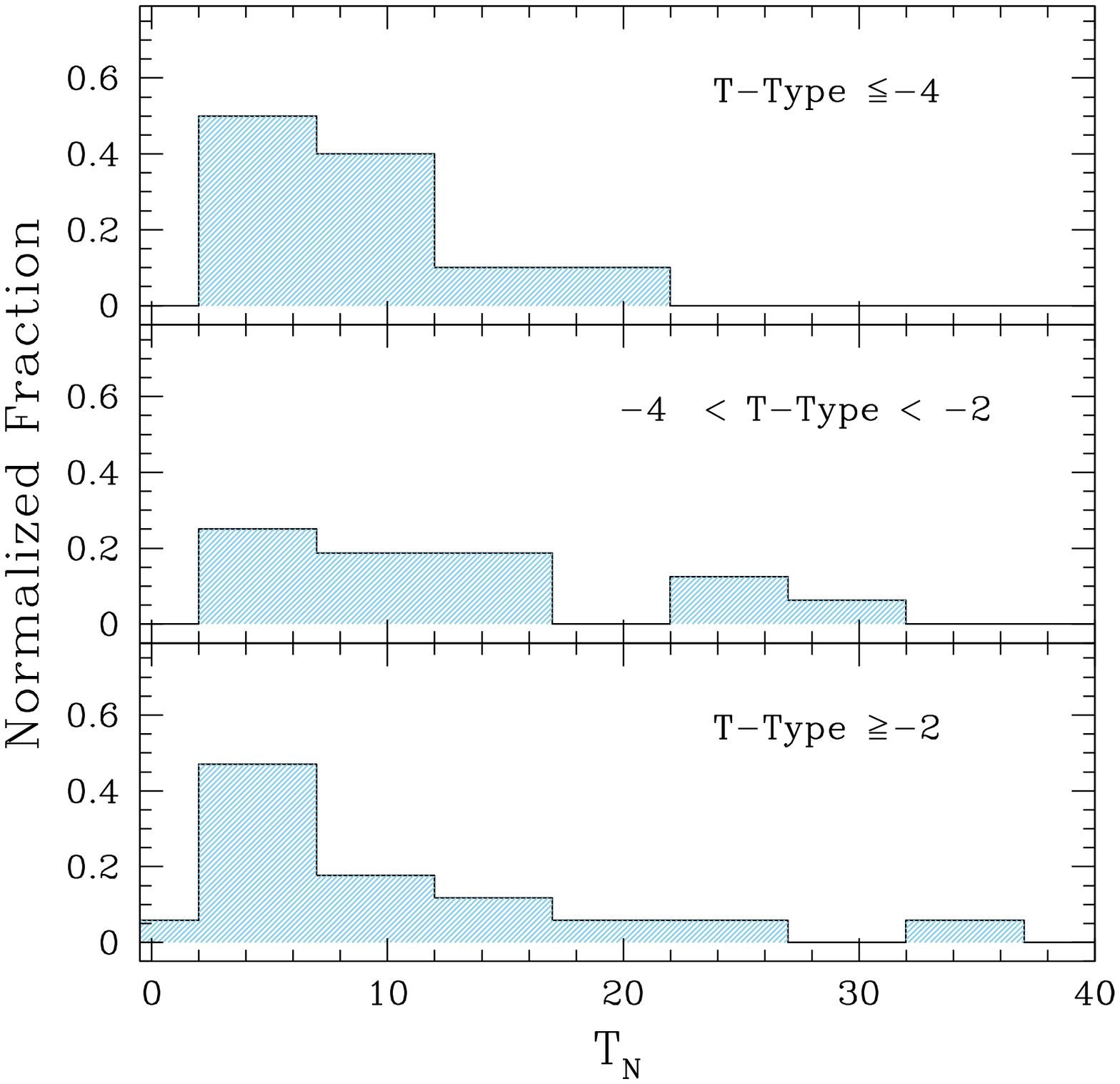}
\caption{Globular cluster specific frequency, T$_{\rm N}$, plotted  for three different sets of parent galaxy morphologies for galaxies with M$_* > 10^{10}$ M$_\odot$. All three classes of galaxies have similar peaks in the distribution of T$_{\rm N}$ at small values and long tails that decline quickly after T$_{\rm N}$ reaches a few.}
\label{fig:sn}
\end{figure}

Finally, morphology, at least over this limited range of early type galaxies, appears to play at most a minor role in determining T$_{\rm N}$. In Figure \ref{fig:sn} we compare the distribution of T$_{\rm N}$ for three different subtypes of galaxies. No strong difference stands out, with all three types having predominantly more galaxies with low T$_{\rm N}$, and all having at least a few galaxies with moderate T$_{\rm N}$. Larger samples will be needed to tease out the hints of differences that appear among the panels if they exist, but such differences would complicate the analysis. For example, \cite{spitler08b} found stronger correlation between T$_{\rm N}$ and M$_*$ than what we find, but they include spirals in their sample. Judging from their Figure 15, the inclusion of spirals, which appear as a class to have lower T$_{\rm N}$ than the earlier types, help drive the correlations because they also tend to have lower M$_*$. It will therefore be imperative in future studies that include a broad range of morphological types to allow for a dependence of T$_{\rm N}$ on both M$_*$ and morphology.

\subsection{The Effect of IMF Variations}

When considering the behavior of N$_{\rm CL}$ with M$_*$, 
we have exclusively concerned ourselves with uncertainties in the ordinate of Figure \ref{fig:numbers} so far. However, the stellar masses are also vulnerable. Although there are systematic uncertainties in the overall mass normalization due to uncertainty in the true IMF \citep[see][for discussion]{meidt14}, changing the mass normalization from that advocated by \cite{eskew} simply shifts all the points the same distance along the abscissa and leaves the N$_{\rm CL} - {\rm M}_*$ slope unchanged. However, if the IMF variation is correlated to 
a galaxy's total stellar mass, such behavior will alter the slope of the N$_{\rm CL}-{\rm M}_*$ relationship. All of the existing mass estimators \citep[such as that from][]{bell,eskew,meidt14} are predicated on a universal IMF, so no existing simple mass estimator resolves this issue.

We approach the problem in an orthogonal manner, by postulating that the cluster specific frequency should not depend on M$_*$. Although there is no known reason why this must be the case, it is a natural possibility. Because our current data indicate that it depends weakly on M$_*$ (see Figure \ref{fig:numbers} and previous discussion), we now ask what IMF behavior would completely remove this trend. If the true stellar mass is M$^\prime_*$, and M$_*$ is the mass inferred using a Salpeter IMF, M$_{Salpeter}$, which is the case for the \cite{eskew} estimator that we use, then $\log ({\rm M}^\prime_*/{\rm M}_{Salpeter}) \propto 0.39 \log({\rm M}_{Salpeter}/{\rm M}_\odot)$ (where the measured M$_* \equiv$ M$_{Salpeter}$) is required to set the slope of the relationship between log N$_{\rm CL}$ and $\log({\rm M}_*/{\rm M}_\odot)$ to 1 (using our previous fit to the relation between these quantities for galaxies with log(M$_*/{\rm M}_\odot ) > 10$).

\begin{figure}[t]
\plotone{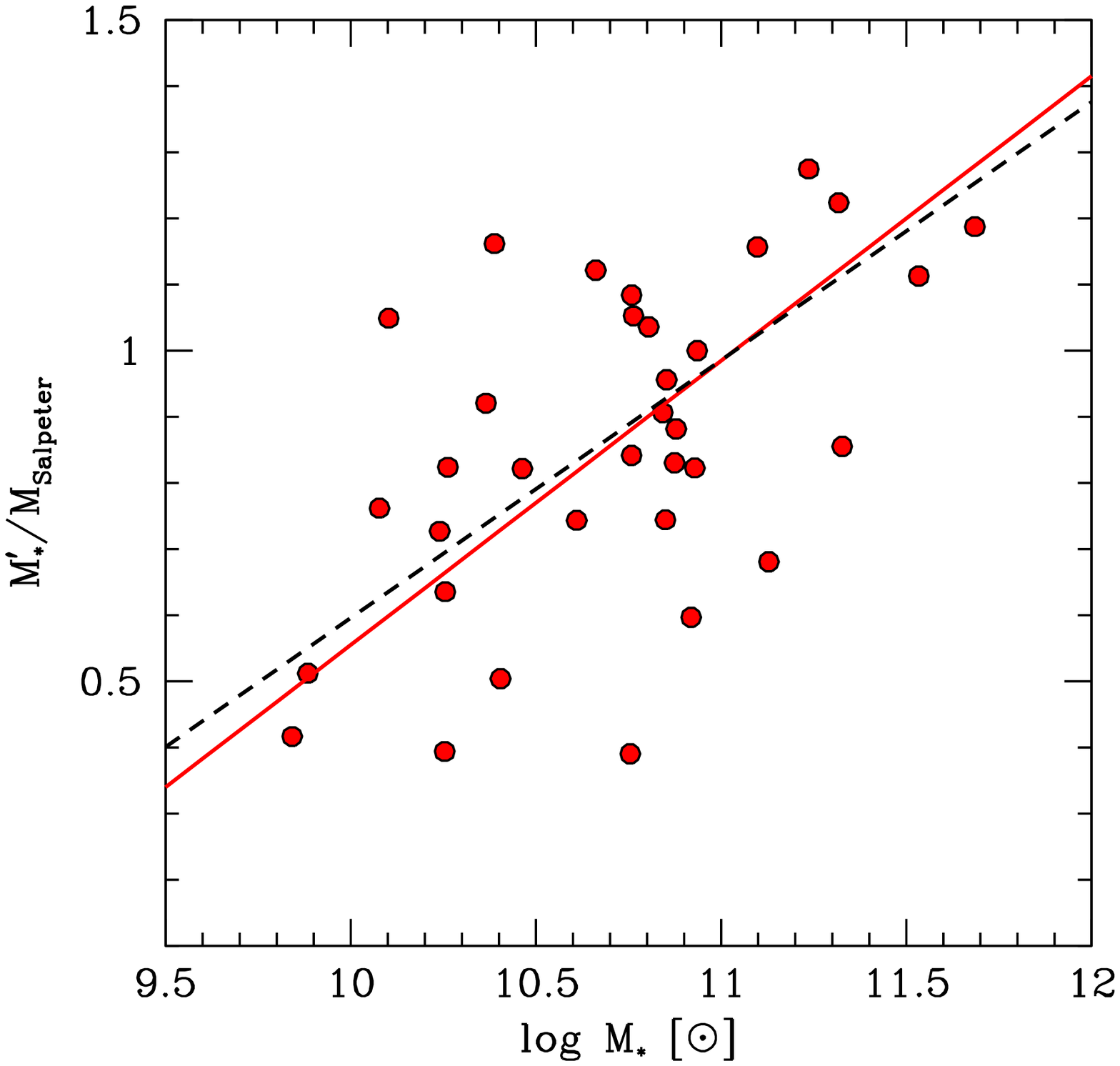}
\caption{Tracing IMF variations. Using the data from the study of \cite{conroy}, we plot their data for the ratio between the actual stellar mass, M$^\prime_*$, and that derived assuming a universal Salpeter IMF against the galaxy's stellar mass. Their stellar mass estimates are derived through a detailed analysis of the integrated spectra of nearby early type galaxies (we exclude the bulge of M 31 from their sample) and stellar population models. The black dotted line is our orthogonal regression fit to these data. The solid red line is not a fit to data shown here, but rather the inferred relationship necessary among these quantities to result in a constant T$_{\rm N}$ for galaxies in our S$^4$G data with log(M$_*/{\rm M}_\odot) > 10$, normalized to produce the best match. The agreement between the slopes of the dotted and solid lines is well within 1$\sigma$ in slope ($0.37\pm0.08$ vs. $0.39 \pm 0.36$, respectively).}
\label{fig:imf}
\end{figure} 

In Figure \ref{fig:imf} we show the relationship between M$_*$, M$_*^{\prime}$, and M$_{Salpeter}$ as measured by \cite{conroy} for their sample of galaxies on the basis of detailed analysis of integrated spectra. We evaluate M$_*$ using the I-band luminosities of these galaxies and their tabulated values of M/L$_{\rm I}$. To obtain the ratio of the true stellar mass to that derived with a Salpeter IMF for each galaxy, we use their tabulated values of M/L$_K$ for each galaxy and the M/L$_K$ value for the Milky Way (they use the ratio of these quantities to normalize out age and metallicity differences among their sample galaxies). We then correct this ratio by 1.6, to account for the fact that the MW galaxy in their modeling has a Kroupa IMF rather than a Salpeter IMF, however this simply results in a renormalization of the coordinate axis in Figure \ref{fig:imf}. Lastly, in the same Figure we plot the relationship discussed previously,
 $\log ({\rm M}^\prime_*/{\rm M}_{Salpeter}) \propto 0.39 \log ({\rm M}_{Salpeter}/{\rm M}_\odot)$, normalized to best match the \cite{conroy} data. 
 
 The quantitatively excellent match of the observed trend in the \cite{conroy} sample and our deduced IMF behavior from the proposition that T$_{\rm N}$ is independent of M$_*$ helps bolster the arguments on both sides --- that the IMF variations are indeed real and that T$_{\rm N}$ is independent of M$_*$. To be precise, the result of the orthogonal regression for the \cite{conroy} sample shown in the Figure (again using the formulas from \cite{isobe}) has a slope of $0.37 \pm 0.08$, well within 1$\sigma$ of our slope estimate of 0.39 derived from the proposition that T$_{\rm N} =$ constant.
 
We pause briefly here to stress that the results we present are only a weak indirect argument for IMF variations. A dependence of T$_{\rm N}$ on stellar mass could arise from a variety of physical factors or even from systematic errors in our estimates of N$_{\rm CL}$. Our principal point here is that the magnitude of such effects is consistent with the magnitude of the effect arising from IMF variations that are suggested by completely different lines of evidence. As such, IMF variations cannot be ignored when considering trends in T$_{\rm N}$.
 
 The most pertinent physical measurement of the specific frequency is perhaps the ratio of mass in globular clusters to the total stellar mass of the galaxy (rather than the {\sl number} of clusters). This approach has been advocated by \cite{mclaughlin} and \cite{harris13}, and that ratio is defined as $\epsilon^b$ (they also advocate using the baryonic mass rather than the stellar mass, and perhaps even relating this all to the halo mass). In the standard approach of a universal IMF and cluster luminosity function, this distinction would not result in any difference in what we have discussed prior to this section. Even if the IMF varies, but it varies in concert for both the stellar populations within galaxies and their globular clusters, then a specific frequency defined as the ratio of the masses of clusters to total stellar mass would be independent of the adopted IMF and any variations of that IMF with M$_*$. 
 
In general, this type of discussion assumes that the IMF within the clusters themselves is universal. There is, however, evidence that the IMF among clusters can also vary \citep{z12,z13}, although the driver of that variation has not yet been identified, so we cannot attempt to model its potential impact on the results described here.
Such variations would affect analyses that are based on the stellar mass within clusters \citep[see][for a discussion relating stellar mass in clusters to galaxy total mass]{harris13}. 

These subtleties become important when trying to place the results in the more global context of galaxy formation. For example, the massive (log(M$_*/{\rm M}_\odot) >$ 10) portion of the sample shown in Figure \ref{fig:specfreq} has a mean value of T$_{\rm N}$ of 11.8 that when combined with the mean cluster mass of $1.2\times 10^5$ M$_{\odot}$ obtained from our adopted LF and the \cite{eskew} mass calibration, suggests a cluster formation efficiency by stellar mass of 0.14\% for an assumed universal IMF. Uncertainties of a factor of two in these assumptions are well within our expectations. \cite{mclaughlin} measures $\epsilon^b = 0.26 \pm 0.05$\%, so about a factor of two larger than our estimate. Whether this difference is due to differences in the galaxy sample, differences in stellar mass estimates, cluster counts, or simple statistical noise (see Figure \ref{fig:specfreq}) is not evident. As such, uncertainties of this magnitude impact arguments relating to whether the cluster population scales more directly to stellar or total mass \citep[eg.][]{strader,georgiev,harris13}.

Interestingly, the high efficiency suggested by the richest low mass systems, about a factor of 10 larger than that for the higher mass galaxies, is comparable to the efficiency derived for cluster formation in dwarf galaxies at high redshift \citep{elmegreen12}. This agreement suggests that the cluster populations in these galaxies may be undisturbed,  with little cluster destruction beyond cluster ``infant mortality".  In more massive galaxies, cluster destruction is expected to be significant \citep{gnedin}, with destruction fractions that can be as large as 90\% (allowing for a reconciliation of our efficiencies in low and high mass galaxies). Unfortunately, any quantitative prediction of the destroyed cluster fraction depends on various assumptions, including the characteristics of the original population.

\section{Summary}

The S$^4$G images \citep{sheth} provide another opportunity to explore the properties of globular cluster populations in galaxies. In particular, the images are sufficiently deep to reach well into the globular cluster luminosity function and have large fields of view, enabling reasonably complete surveys for clusters in a sample of thousands of galaxies that surveys the local volume. Although, the data provide neither the angular resolution nor multi wavelength coverage of the latest best examples of similar efforts, the strength of this program is in its large, homogeneous sample of galaxies. The results obtained here on the basic properties of globular cluster populations, number and specific frequency as a function of various galaxy properties, complement the higher fidelity surveys undertaken with {\sl HST} and deep ground based surveys. Our empirical findings are consistent with those based on compilations of existing data \citep[eg.][]{harris13}, but given the nature of the literature samples, where different categories of galaxies are the focus of independent studies, we advocate that quantitative measures and comparison be based on homogeneous data and analysis.

We present our initial foray into this topic with S$^4$G data, concentrating here on early type galaxies in the core survey. This subsample consists of 97 galaxies drawn from the over 2300 galaxies in the S$^4$G dataset. We avoided late-type galaxies to minimize confusion between H{\small II} regions, luminous stars, and globular clusters, but an extension of this work to edge-on late type galaxies should be straightforward. A large sample of well-studied edge-on galaxies exists within S$^4$G \citep{comeron}. In addition, we have begun a program to enlarge the complement of early-type galaxies in S$^4$G, galaxies that our initial selection criteria discriminated against, and so an extension of this work to include the larger sample of early type galaxies will  be forthcoming.

Using this initial set of galaxies we established the following:

$\bullet$ Accurate measurements of the number of clusters, N$_{\rm CL}$,  in each galaxy can best be obtained by fitting a radial surface density profile to all detected point sources with absolute magnitudes between $-11$ and $-8$, if sources are assumed to be at the distance of the galaxy, that is a power law with an exponent of $-2.4$. The normalization and background level (i.e. surface density of contaminants) are fit. Comparison to independent measurements of N$_{\rm CL}$ demonstrates that our measurements have an uncertainty of a factor of $\sim$ 2 and are well described by the internal uncertainties we calculate.

$\bullet$ Over the full range of galaxy stellar masses, M$_*$, N$_{\rm CL}$ rises nearly proportionally to M$_{*}$ (log(N$_{\rm CL}) \propto (0.94\pm0.19) \log({\rm M}_*/{\rm M}_\odot)$). However, the behavior of N$_{\rm CL}$ appears to change significantly below log(M$_*/{\rm M}_\odot) \sim 10$. This result confirms previous findings \citep{georgiev,harris13}. If we only fit to galaxies with log(M$_*/{\rm M}_\odot) > 10$, then the relationship steepens to log(N$_{\rm CL}) \propto (1.39\pm0.36) \log({\rm M}_*/{\rm M}_\odot)$. Because of the large uncertainty, we have only marginal evidence that this slope is larger than 1 (although a bisector fit of the data results in a similar result, slope $= 1.24\pm0.19$). Definitive evidence of a slope larger than 1 would demonstrate that more massive galaxies, or their progenitors, are somehow more efficient producers of globular clusters and so cannot simply be the result of dissipationless mergers of lower mass early-type galaxies. 

$\bullet$ Because determining if T$_{\rm N}$ varies with galaxy stellar mass is critical for cluster formation models, we took care to quantify the effect of a varying stellar initial mass function (IMF) on the determination of the galaxy stellar mass. In particular, we demonstrate that if the IMF of these galaxies depends on galaxy mass \citep[as has been suggested by a number of recent studies;][]{conroy,cappellari}, the apparent, but weak, increase in T$_{\rm N}$ with M$_*$ is entirely removed. In fact, when we postulate that T$_{\rm N}$ is constant with M$_*$, we recover the exact dependence of the IMF with stellar mass observed in the data of \cite{conroy}. We conclude that T$_{\rm N}$ is independent of mass for early-type galaxies with log (M$_*/{\rm M}_\odot) > 10$. It is possible that the most massive ellipticals, or central dominant ellipticals deviate from this trend \citep{mclaughlin,peng08}, we have insufficient data to constrain T$_{\rm N}$ for that subclass. However, the lack of a dependence of T$_{\rm N}$ on M$_*$ in general is consistent with simple predictions from models of dissipationless mergers and of dissipational mergers where any increased star formation during the merger stage produces a corresponding number of globular clusters. With respect to the latter scenario, we note that globular cluster candidates consistent in age with the star burst phase in E+A galaxies have been detected \citep{yang}. 

$\bullet$ At log(M$_*/{\rm M}_\odot) < 10$ the scatter in N$_{\rm CL}$ is large (2 orders of magnitude) and so may suggest that we are detecting strong differences in cluster formation  and destruction efficiencies among this population of galaxies.  This result confirms previous studies of the specific frequency among low mass galaxies \citep{miller,strader,georgiev}. Our data also hint at the possibility of two T$_{\rm N}$ families of galaxies at low M$_*$, also supporting previous speculations \citep{miller,strader}. Progress in understanding the drivers of cluster formation may come principally from a focused effort in studying the cluster populations of low mass galaxies. However, because of these populations are inherently poor in numbers, large samples will be needed to counter the larger statistical uncertainties associated with the cluster populations of individual galaxies. 

Using the S$4$G images, and the larger sample of early-type galaxies currently being collected, we expect this work to expand to a significantly larger sample of galaxies that both produces more statistically significant results for the cluster populations of early-types and a comparison sample of the cluster populations of late-type galaxies. Ultimately, the power of having measurements of T$_{\rm N}$ for the S$^4$G sample will also lie in the detailed ancillary analysis products such as photometric decomposition, detailed morphologies, radial and vertical disk studies, and measurements of stellar and gas masses. In all, we expect this series of papers to provide a broad but comprehensive outline of the basic properties of globular clusters systems in galaxies.

\begin{acknowledgments}

DZ acknowledges financial support from 
NASA ADAP NNX12AE27G and  NYU CCPP for its hospitality during long-term visits. 
LCH acknowledges support from the Kavli Foundation, Peking University, and the Chinese Academy of Science through grant No. XDB09030102 (Emergence of Cosmological Structures) from the Strategic Priority Research Program.  EA and AB acknowledge financial support from the CNES
(Centre National d'Etudes Spatiales - France) and from the People
Programme (Marie Curie Actions) of the European Union's Seventh
Framework Programme FP7/2007-2013/ under REA grant agreement number
PITN-GA-2011-289313 to the DAGAL network. The authors thank Ryan Leaman and Mike Beasley for valuable comments.
This research has made use of the NASA/IPAC Extragalactic Database (NED), which is operated by the Jet Propulsion Laboratory, California Institute of Technology, under contract with NASA. 
The National Radio Astronomy Observatory is a facility of the National Science Foundation operated under cooperative agreement by Associated Universities, Inc. Finally, we thank our conscientious referee for valued comments and suggestions.
\end{acknowledgments}

\begin{deluxetable*}{lrrrrrrrrrrrr}
\tablecaption{Globular Cluster Population Properties}
\tablewidth{0pt}
\tablehead{
\colhead{Name}&
\colhead{DM}&
\colhead{T}&
\colhead{m$_{3.6}$}&
\colhead{m$_{4.5}$}&
\colhead{N$_{50}$}&
\colhead{T$_{\rm N}$}&
\colhead{Back.}&
\colhead{a}&
\colhead{b}&
\colhead{Q}\\
}
\startdata
ESO 357-025&  31.13 &  $-$2 &    14.9 &    15.4 &       20 &     6.59$_{-4.86}^{+6.16}$&   $-$5.38 &    12.41 &    $-$5.84 &   1\\
ESO 419-013&  31.45 &  $-$2 &    14.5 &    14.8 &        98 &   16.34$_{-11.32}^{+11.14}$ & $-$5.00 &    3.91 &    $-$2.98 &    1\\
ESO 548-023&  31.71 &  $-$3 &    14.4 &    14.8 &       111&     13.44$_{-12.49}^{+15.01}$ &  $-$5.35 &    14.96 &    $-$5.25 &      0\\
IC 335 &  31.19 &  $-$3 &    11.9 &    12.4 &       2&     0.04$_{-0.01}^{+0.86}$ &  $-$5.49 &     0.01 &    $-$1.45 &     0\\
IC 796 &  32.06 &  $-$3 &    13.0 &    13.5 &        153 &     3.67$_{-1.24}^{+1.75}$ &    $-$4.80 &    11.71 &    $-$5.20 &   1\\
\enddata
\label{tab:results}
\tablecomments{
DM refers to the distance modulus. T is the morphological T$-$Type of the galaxy from the compilation of \cite{buta14}. The 3.6 and 4.6$\mu$m magnitudes of the galaxies as measured by \cite{munoz} are in columns 4 and 5. N$_{50}$ is the number of globular clusters estimated from our best fit model of fixed power-law slope within 50 kpc of the galaxy. T$_{\rm N}$ is the number of these clusters per 10$^9$ M$_\odot$ of stellar mass in the galaxy. The quantity Back. represents the log of the measured background level (sources/pc), and a and b represent the normalization and slope of the power law fit, for models where the power-law if allowed to float and the background is set to Back.
This table is available in its entirety in a machine-readable form in the online journal. A portion is shown here for guidance regarding its form and content.}
\end{deluxetable*}

\end{document}